\documentclass[prl,twocolumn,superscriptaddress]{revtex4-1}
\usepackage{}
\usepackage{amssymb}
\usepackage{amsfonts}
\usepackage{graphics,graphicx,epsfig,bm,amsmath,amsthm,amssymb}
\usepackage{bm}
\usepackage{bbm}
\usepackage{longtable}
\usepackage{multirow}
\usepackage{array}
\usepackage{color}
\usepackage[usenames,dvipsnames]{xcolor}

\usepackage{float}
\usepackage[a4paper,colorlinks=true,
linkcolor=blue,citecolor=blue,
pdfauthor={ },
pdftitle={ },
pdfsubject={ },
pdfkeywords={ }]{hyperref}

\begin{document}

\title{Floquet dynamical quantum phase transitions}
%Observation of Interband Coherence Effect in Thouless Pump on a Single Spin in Diamond

\author{Kai Yang}
\affiliation{Hefei National Laboratory for Physical Sciences at the Microscale and Department of Modern Physics, University of Science and Technology of China, Hefei 230026, China}
\affiliation{CAS Key Laboratory of Microscale Magnetic Resonance, University of Science and Technology of China, Hefei 230026, China}
%\affiliation{Synergetic Innovation Center of Quantum Information and Quantum Physics, University of Science and Technology of China, Hefei 230026, China}

\author{Longwen Zhou}
\affiliation{Department of Physics, College of Information Science and Engineering, Ocean University of China, Qingdao 266100, China}
%\affiliation{Department of Physics, National University of Singapore, Singapore 117543}

\author{Wenchao Ma}
\affiliation{Hefei National Laboratory for Physical Sciences at the Microscale and Department of Modern Physics, University of Science and Technology of China, Hefei 230026, China}
\affiliation{CAS Key Laboratory of Microscale Magnetic Resonance, University of Science and Technology of China, Hefei 230026, China}
%\affiliation{Synergetic Innovation Center of Quantum Information and Quantum Physics, University of Science and Technology of China, Hefei 230026, China}

\author{Xi Kong}
\affiliation{The State Key Laboratory of Solid State Microstructures and Department of Physics, Nanjing University, Nanjing 210093, China}

\author{Pengfei Wang}
\affiliation{Hefei National Laboratory for Physical Sciences at the Microscale and Department of Modern Physics, University of Science and Technology of China, Hefei 230026, China}
\affiliation{CAS Key Laboratory of Microscale Magnetic Resonance, University of Science and Technology of China, Hefei 230026, China}
\affiliation{Synergetic Innovation Center of Quantum Information and Quantum Physics, University of Science and Technology of China, Hefei 230026, China}

\author{Xi Qin}
\affiliation{Hefei National Laboratory for Physical Sciences at the Microscale and Department of Modern Physics, University of Science and Technology of China, Hefei 230026, China}
\affiliation{CAS Key Laboratory of Microscale Magnetic Resonance, University of Science and Technology of China, Hefei 230026, China}
\affiliation{Synergetic Innovation Center of Quantum Information and Quantum Physics, University of Science and Technology of China, Hefei 230026, China}

\author{Xing Rong}
\affiliation{Hefei National Laboratory for Physical Sciences at the Microscale and Department of Modern Physics, University of Science and Technology of China, Hefei 230026, China}
\affiliation{CAS Key Laboratory of Microscale Magnetic Resonance, University of Science and Technology of China, Hefei 230026, China}
\affiliation{Synergetic Innovation Center of Quantum Information and Quantum Physics, University of Science and Technology of China, Hefei 230026, China}

\author{Ya Wang}
\affiliation{Hefei National Laboratory for Physical Sciences at the Microscale and Department of Modern Physics, University of Science and Technology of China, Hefei 230026, China}
\affiliation{CAS Key Laboratory of Microscale Magnetic Resonance, University of Science and Technology of China, Hefei 230026, China}
\affiliation{Synergetic Innovation Center of Quantum Information and Quantum Physics, University of Science and Technology of China, Hefei 230026, China}

\author{Fazhan Shi}
\email{fzshi@ustc.edu.cn}
\affiliation{Hefei National Laboratory for Physical Sciences at the Microscale and Department of Modern Physics, University of Science and Technology of China, Hefei 230026, China}
\affiliation{CAS Key Laboratory of Microscale Magnetic Resonance, University of Science and Technology of China, Hefei 230026, China}
\affiliation{Synergetic Innovation Center of Quantum Information and Quantum Physics, University of Science and Technology of China, Hefei 230026, China}

\author{Jiangbin Gong}
\email{phygj@nus.edu.sg}
\affiliation{Department of Physics, National University of Singapore, Singapore 117543}

\author{Jiangfeng Du}
\email{djf@ustc.edu.cn}
\affiliation{Hefei National Laboratory for Physical Sciences at the Microscale and Department of Modern Physics, University of Science and Technology of China, Hefei 230026, China}
\affiliation{CAS Key Laboratory of Microscale Magnetic Resonance, University of Science and Technology of China, Hefei 230026, China}
\affiliation{Synergetic Innovation Center of Quantum Information and Quantum Physics, University of Science and Technology of China, Hefei 230026, China}

%\pacs{76.30.Mi (color centers), 76.70.Hb (ODMR), 07.55.Ge (Magnetometry)}

\begin{abstract}
Dynamical quantum phase transitions (DQPTs) are manifested by time-domain nonanalytic behaviors of many-body systems.
Introducing a quench is so far understood as a typical scenario to induce DQPTs.
In this work, we discover a novel type of DQPTs, termed ``Floquet DQPTs", as intrinsic features of systems with periodic time modulation.
Floquet DQPTs occur within each period of continuous driving, without the need for any quenches.
In particular, in a harmonically driven spin chain model, we find analytically the existence of Floquet DQPTs in and only in a parameter regime
hosting a certain nontrivial Floquet topological phase. The Floquet DQPTs are further characterized by a dynamical topological invariant defined as the winding number of the Pancharatnam geometric phase versus quasimomentum.  These findings are experimentally demonstrated with a single spin in diamond.
This work thus opens a door for future studies of DQPTs in connection with topological matter.
\end{abstract}
\maketitle

%DQPTs are characterized by time-domain non-analyticities of certain observables in many-body systems \cite{HeylPRL2013}.
%Their occurrence is often induced by a quench across an equilibrium
%quantum critical point \cite{PollmannPRE2010}.
DQPTs are often associated with quantum quenches, a protocol in which parameters of a Hamiltonian are suddenly changed~\cite{HeylPRL2013,PollmannPRE2010}.
%The quantum quench is a protocol in which parameters of a Hamiltonian are suddenly changed \cite{Mitra}.
A quantum quench across an equilibrium quantum critical point may induce a DQPT.
%Typically, the occurrence of DQPT a quantum quench across an equilibrium quantum critical point
%The occurrence of DQPTs is often induced by a quantum quench across an equilibrium quantum critical point.
If pre-quench and post-quench systems are in
topologically distinct phases, DQPTs may also be characterized
by dynamical topological invariants~\cite{DTOP1,DTOP2,DTOP3}.  As a promising
%an attractive and fruitful
approach to classify quantum states of matter in nonequilibrium
situations, DQPTs have been theoretically explored in both closed
%\cite{KarraschPRB2013,HeylPRL2014,CanoviPRL2014,VajnaPRB2014,HickeyPRB2014,AndraschkoPRB2014,KrielPRB2014,HeylPRL2015,SharmaPRB2015,HuangPRL2016,AbelingPRB2016,SharmaPRB2016,PuskarovSPP2016,BhattacharyaPRB2017,HeylPRB2017,KarraschPRB2017,HomrighausenPRB2017,HalimehPRB2017,Zauner-StauberPRE2017,FogartyNJP2017,TrapinPRB2018,BhattacharjeePRB2018,SedlmayrPRB2018-2,GurarieArXiv2018,LyuArXiv2018}
and open quantum systems
% \cite{ZhouArXiv2017,HeylPRB2017-2,BhattacharyaPRB2017-3,LangPRB2018,SedlmayrPRB2018,LangArXiv2018,QiuArXiv2018,BandyopadhyayArXiv2018}
at different physical dimensions~\cite{DQPTRev1,DQPTRev2,ZhouArXiv2017}.
% \cite{SchmittPRB2015,BhattacharyaPRB2017-2,WangPRA2018,SchmittSSP2018}.
Experimentally, DQPTs have been observed in trapped ions \cite{DQPTExp1,DQPTExp2},
cold atoms \cite{DQPTExp3,DQPTExp4}, superconducting qubits \cite{DQPTExp5},
nanomechanical oscillators \cite{DQPTExp6}, and photonic quantum walks \cite{DQPTExp7,DQPTExp8}.

\begin{figure}
	\includegraphics[width=1\columnwidth]{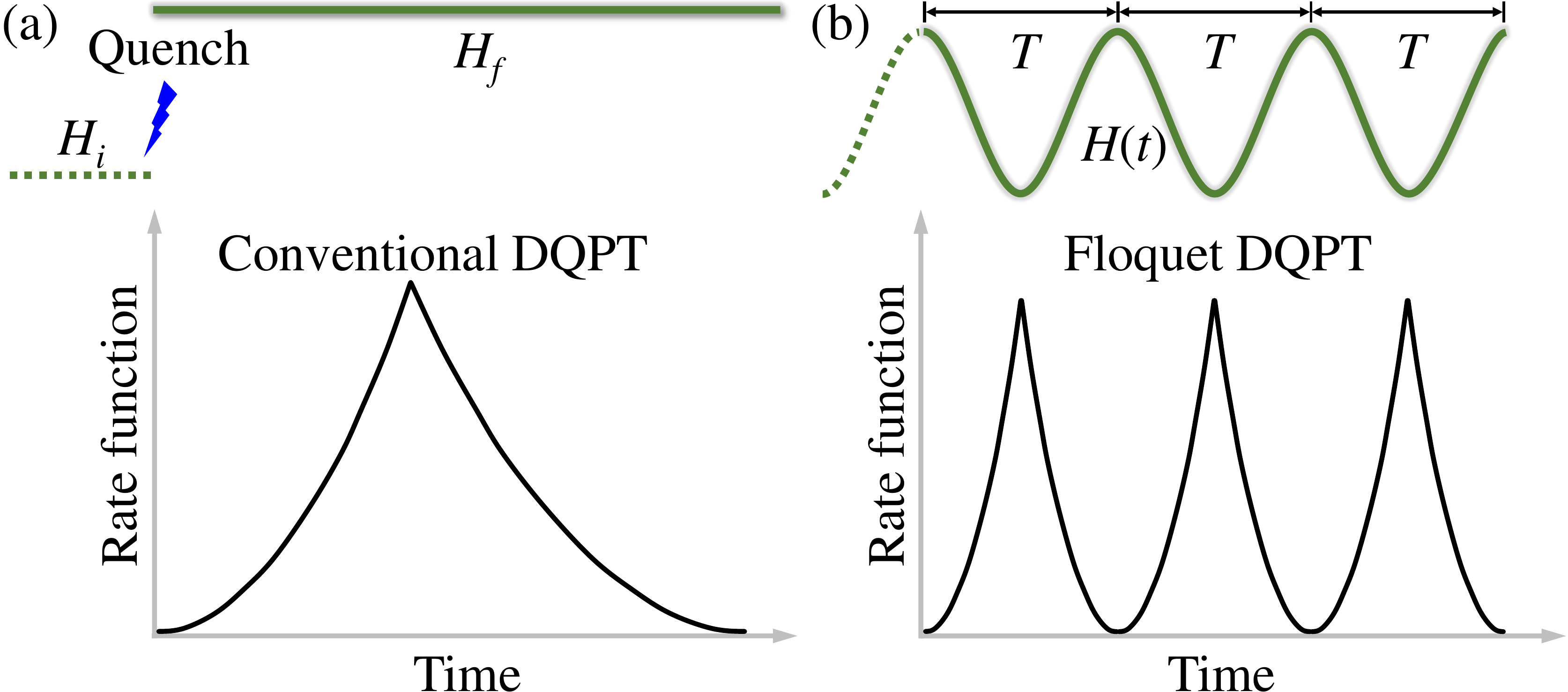}
	\caption{Comparison between (a) DQPTs following a quantum quench, and (b) Floquet DQPTs without any quenches.
		  Here $H_i$ and $H_f$ denote the Hamiltonians before and after the quench, $H(t)$ denotes a periodically and continuously modulated Hamiltonian.
	}
	\label{theory}
\end{figure}

To date, in most studies of DQPTs,
%theoretical studies and all experimental studies of DQPTs,
a quantum quench acts as a trigger for initiating nonequilibrium dynamics and then exposing the underlying topological features.
%which is deemed rather necessary to initiate nonequilibrium dynamics and then to reveal the underlying topological features.
However, DQPTs under more general nonequilibrium manipulations are still largely unexplored \cite{SharmaPRB2016,PuskarovSPP2016,BhattacharyaPRB2017}.
In particular, because the dynamics of systems under time-periodic modulations has led to fascinating discoveries like Floquet topological states~\cite{Lindner2011,FTIRev,LW2014,HoPRL2012,LiPRL2018} and
discrete time crystals \cite{TCRev,Yao2018,BomantaraPRL}, it is urgent to investigate how DQPTs may occur in such Floquet systems.
%rich in topological features.
%\cite{DQPTExp7,DQPTExp8,FDQPT1,FDQPT2},
Along this avenue, there have been scattered studies, but still with the notion that DQPTs are best aroused by a quench to some system parameters~\cite{FDQPT1,FDQPT2}.  Here we introduce a novel class of DQPTs, termed Floquet DQPTs,  which can be regarded as
intrinsic features of systems with time-periodic modulations.
As schematically shown in Fig.~\ref{theory}, the Floquet DQPTs we discovered
%in this work
occur within each period of a continuous driving field, without the need for any quenches.
In the model investigated below, the occurrence of Floquet DQPTs and the emergence of Floquet topological phases are in the same parameter regime.
We further perform a quantum simulation experiment using a single electron spin in diamond to verify our main theoretical findings.

%\emph{Theory}.---
For a periodically modulated system ${\cal H}(t)={\cal H}(t+T)$ with $T$  the modulation period and  $\omega=2\pi/T$  the frequency,
 the time-dependent Schr\"{o}dinger equation yields solutions  $|\Psi_{n}(t)\rangle=e^{-i\varepsilon_{n}t}|\Phi_{n}(t)\rangle$, with the Floquet mode $|\Phi_{n}(t)\rangle=|\Phi_{n}(t+T)\rangle$ and the quasienergy $\varepsilon_{n}\in[-\omega/2,\omega/2)$~\cite{HolthausFloTut}.
%A complete set of solutions of the time-dependent Schr\"{o}dinger equation is given by the so-called Floquet states $\{|\psi_{n}(t)\rangle\}$.
%Each Floquet state can be decomposed as $|\psi_{n}(t)\rangle=e^{-i\varepsilon_{n}t}|\varphi_{n}(t)\rangle$,
%with the Floquet mode $|\varphi_{n}(t)\rangle=|\varphi_{n}(t+T)\rangle$ and the quasienergy $\varepsilon_{n}\in[-\omega/2,\omega/2)$.
%The time evolution operator can be expanded by the Floquet modes as
%${\cal U}(t)=\sum_{n}e^{-i\varepsilon_{n}t}P(t)|\Phi_{n}(0)\rangle\langle\Phi_{n}(0)|$,
%where $P(t)=P(t+T)$ is the micromotion operator~\cite{SM}.
Though a Floquet mode by definition becomes parallel to itself at multiple $T$'s,
it still evolves nontrivially within each driving period (hence the micromotion dynamics).
%As we shall see below,  non-analyticities can emerge from the micromotion dynamics,
%and we define such behavior as Floquet DQPTs.
More specifically, consider a periodically driven system prepared initially
at a many-particle Floquet state $|\Psi_n(0)\rangle$, which is the product state of Floquet modes $|\varphi_{n}(0)\rangle$ filling the quasienergy band $n$.
%The return amplitude of the system at $t>0$ is given by $G(t)=e^{-i\varepsilon_{n}t}\langle\phi_{n}(0)|\phi_{n}(t)\rangle$.
The return amplitude of the system at $t>0$ is then ${\cal G}_n(t)\equiv\langle\Psi_{n}(0)|\Psi_{n}(t)\rangle$.
In the thermodynamic limit, the rate function is defined as
\begin{equation}
g_n(t)\equiv-\lim_{N\rightarrow\infty}\frac{1}{N}\ln|{\cal G}_n(t)|^{2},\label{eq:RateFuc}
\end{equation}
where $|{\cal G}_n(t)|^{2}$ is the return probability, and $N$ is the number of degrees of freedom, such as the number of lattice sites in a chain.
The rate function $g_n(t)$ plays the role of a dynamical free energy~\cite{HeylPRL2013,SM}, which is periodic in time with period $T$.
Now if there exists a critical time $t_{c}$, at which $|\Psi_{n}(t_{c})\rangle$ is orthogonal to $|\Psi_{n}(0)\rangle$, one finds ${\cal G}_{n}(t_{c})=0$.
Then $g_n(t)$ as defined in Eq.~(\ref{eq:RateFuc}) or its time derivatives becomes nonanalytic at $t=t_{c}$.
%representing a unique feature of Floquet DQPTs.
%Notably, the occurrence of such DQPTs do not need any quench and
%the periodic driving can be just a simple harmonic modulation to some system parameters.

%However, a thorough theoretical understanding of Floquet DQPTs is
%challenging, since a periodically driven quantum system is usually
%nonintegrable, and perturbative treatments work well only at high
%and low driving frequencies.

%\emph{Model}.---
 We now focus on a harmonically driven spin chain described by
%\begin{widetext}
$H(t) = \sum_{n}[(\delta_{1}-\Omega\sin\omega t)s_{n}^{x}s_{n+1}^{x}+(\delta_{1}+\Omega\sin\omega t)s_{n}^{y}s_{n+1}^{y}-\delta_{2}s_{n}^{z}\nonumber-\Omega\cos\omega t(s_{n}^{x}s_{n+1}^{y}+s_{n}^{y}s_{n+1}^{x})]$,
where $s_{n}^{x,y,z}$ are quantum spin-$1/2$
variables localized at site $n$ of the chain; $\delta_{1}$, $\delta_{2}$, and $\Omega$ are system parameters.
The associated Bloch Hamiltonian is found to be~\cite{SM}
\begin{equation}
H(k,t)=\frac{\Omega\sin k}{2}[\cos(\omega t)\sigma_{x}+\sin(\omega t)\sigma_{y}] + \frac{\delta_{1}\cos k+\delta_{2}}{2}\sigma_{z}.
\end{equation} where $k$ is the quasimomentum.
The dynamics governed by $H(k,t)$ is analytically solvable and has been already experimentally realized in other quantum simulation studies~\cite{GTPExp}.
%As a side note, it is tempting to transform the harmonically driven Hamiltonian $H(k,t)$ to a static Hamiltonian in a frame rotating at the angular frequency $\omega$.
%However, this is a misconception because the driving itself is essential for generating the nontrivial Floquet topological phase with a bulk-edge correspondence~\cite{AsbothPRB2013} predicting topological edge states under open boundary conditions~\cite{SM}.
%\emph{Results}.---
The bulk dynamics of the system can be solved because each occupied $k$ component of the system evolves independently.
The Floquet state at each $k$ is given by $|\psi_{\pm}(k,t)\rangle=e^{-iE_{\pm}(k)t}U_{R}(t)|\varphi_{\pm}(k,0)\rangle$.
Here $E_{\pm}(k)$ are quasienergy dispersions of two Floquet bands with the Floquet modes $|\varphi_{\pm}(k,t)\rangle = U_{R}(t)|\varphi_{\pm}(k,0)\rangle$,
$U_{R}(t)={\rm diag}(1,e^{i\omega t})$ is responsible for the micromotion dynamics, and
$|\varphi_{\pm}(k,0)\rangle$ are the eigenstates of the static Hamiltonian $H_F(k) = \Omega\sin k ~\sigma_{x}/2 + (\delta_{1}\cos k+\delta_{2}-\omega)\sigma_{z}/2$~\cite{SM}.

%\begin{equation}
%U_{R}(t)=\begin{pmatrix}1 & 0\\
%0 & e^{i\omega t}
%\end{pmatrix},
%\label{Eq:URmain}
%\end{equation}
%and up to a normalization constant,
%\begin{equation}
%|\varphi_{\pm}(k)\rangle=\begin{bmatrix}h_{xy}(k)\\
%E_{\pm}(k)-h_{z}(k)
%\end{bmatrix}.
%\label{Eq:VarphiPMmaintext}
%\end{equation}

The micromotion dynamics arising from $U_{R}(t)$ is the rotation around the $z$ axis (with frequency $\omega$) on the Bloch sphere~\cite{SM}.
If there exists  $k=k_c$ such that $|\varphi_{\pm}(k,0)\rangle$ lies in the $x$-$y$ plane,
then the micromotion dynamics can always rotate this initial state to a state pointing at exactly the opposite direction at $t=t_{c}=(2m-1)T/2$ for $m\in\mathbb{Z}^{+}$.
That is, at $t=t_{c}$ the time-evolving state at this quasimomentum becomes orthogonal to its initial state.
Assuming that the collective initial state of the system fills one of the Floquet bands $|\Psi_{\alpha}(0)\rangle=\prod_{k>0}|\psi_{\alpha}(k,0)\rangle=\prod_{k>0}|\varphi_{\alpha}(k,0)\rangle$ with band indices $\alpha=+$ or $-$.
At a later time $t$, the return amplitude of the system is given by ${\cal G}_{\alpha}(t)=\prod_{k>0} G_{\alpha}(k,t)$, with $G_{\alpha}(k,t)=\langle\psi_{\alpha}(k,0)|\psi_{\alpha}(k,t)\rangle$.
It can be shown that, under the condition
\begin{equation}
\left|\omega-\delta_{2}\right|\leq \left|\delta_{1}\right|,
\label{eq:CriticalMom}
\end{equation}
there exists a critical momentum $k_c$ with $\delta_{1}\cos k_{c}=\omega-\delta_{2}$, such that $|\varphi_{\alpha}(k_c,0)\rangle$ lies in the $x$-$y$ plane~\cite{SM} and consequently $G_{\alpha}(k_{c},t)=0$.
%The existence of $k_c$ here does not require quenching across equilibrium quantum critical points~\cite{HeylPRL2013}.
Under the condition (\ref{eq:CriticalMom}), the rate function $g_{\alpha}(t)=-\int dk\ln|G_{\alpha}(k,t)|^{2}$ becomes non-analytic at each critical time $t_{c}$.
It can be also shown that there exists a topological order parameter given by the winding number of the geometric
phase over the Brillouin zone, namely,
\begin{equation}
{\nu _\alpha }(t) = \int_0^\pi \frac{{dk}}{{2\pi }} \frac{\partial \phi_\alpha^{\rm{geo}}(k)}{\partial k},
\label{eq:winding}
\end{equation}
where $\phi_\alpha ^{{\rm{geo}}}(k)$ is the Pancharatnam geometric phase acquired during the evolution of a Floquet state at quasimomentum $k$~\cite{SM}.
When time passes the critical time of a Floquet DQPT, $\nu _\alpha(t)$ makes a quantized jump.
The winding number $\nu_\alpha(t)$ as a dynamical topological invariant thus reflects the topological nature of Floquet DQPTs identified in the spin chain model here.

%The Floquet operator of the system governed by the Hamiltonian $H(k,t)$ is $U(k,T)={\cal T}e^{-i\int_{0}^{t}dt H(k,t)}$, where ${\cal T}$ denotes the time ordering operator. It can be rotated by unitary transformations to Floquet operators in two complementary symmetric time frames, given by $U_\beta(k)\equiv -e^{-i{\bm d}_\beta(k)\cdot{\bm{\sigma}}}$~\cite{SM}. Here $\beta=1,2$ are the indices of the two time frames, ${\bm d}_\beta(k)=(d_{{\beta}z}(k),d_{{\beta}x}(k))$ and ${\bm \sigma}=(\sigma_z,\sigma_x)$. Since $U_\beta(k)$ has the chiral symmetry $\Gamma=\sigma_y$ in the sense that $\Gamma U_\beta(k)\Gamma=U^{-1}_\beta(k)$, the topological phases of $U(k,T)$ can be characterized by a pair of winding numbers $W_0=(W_1+W_2)/2$ and $W_\pi=(W_1-W_2)/2$, where $W_1$ and $W_2$ are the winding numbers of $U_1(k)$ and $U_2(k)$, defined as
%\begin{equation}
%W_\beta=\int_{-\pi}^\pi\frac{dk}{2\pi}\frac{[{\bm d}_\beta(k)\times\partial_k{\bm d}_\beta(k)]_y}{|{\bm d}_\beta(k)|^2},
%\label{eq:chiral}
%\end{equation}
%for $\beta=1,2$~\cite{AsbothPRB2013}. In the Floquet DQPT regime specified by Eq.~(\ref{eq:CriticalMom}), we have $W_0=0$ and $W_\pi=1$, implying that Floquet states of $U(k,T)$ are topologically nontrivial in this regime~\cite{SM}, with degenerate $\pi$ edge modes existing under open boundary conditions~\cite{AsbothPRB2013}.

Before proceeding to experiment, three traits of Floquet DQPTs deserve to be highlighted.
Firstly, no quantum quench from one equilibrium phase to another is required in our proposal, as the continuous driving field suffices to generate non-analyticities in the time domain.
Secondly, the Floquet DQPTs introduced in this work repeat periodically in time, whereas the DQPTs following a single quench are usually visible only in transient time windows due to the decay of the rate function.
Floquet DQPTs are therefore more accessible to experiments.
Thirdly, precisely under the same condition as in (\ref{eq:CriticalMom}), our spin chain model is found to reside in a nontrivial ``chiral-symmetric" Floquet topological phase~\cite{SM}, as featured by two winding numbers defined with the one-period evolution operator in two chiral-symmetric time frames~\cite{AsbothPRB2013}. These two winding numbers can be used to predict
topological edge states under open boundary conditions.
As such, the Floquet DQPTs reported here not only provide an indirect probe to study these intriguing Floquet topological phases,
but also suggest a remarkable connection between jumps in the dynamical topological invariant ${\nu_\alpha }(t)$ and the emergence of a Floquet topological phase~\cite{SM}.

\begin{figure}
	\includegraphics[width=1\columnwidth]{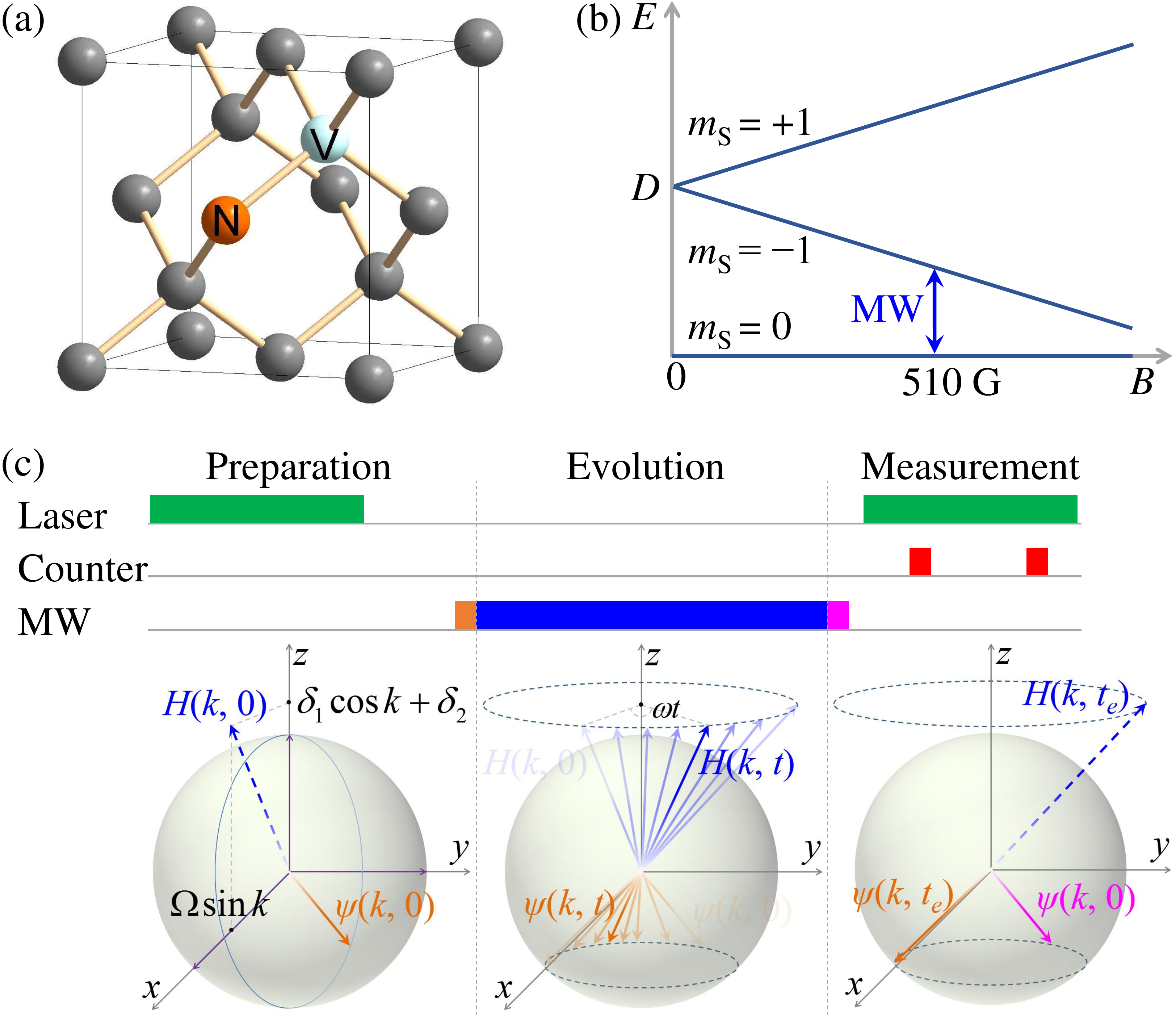}
	\caption{Experimental system and method.
		(a) NV center in diamond. %The NV center consists of a substitutional nitrogen atom and a neighboring vacancy.
		(b) Electronic ground state of a negatively charged NV center. The energy splitting depends on the magnetic field which is parallel to the NV axis in this experiment. The two levels $\left| {{m_s} = 0} \right\rangle$ and $\left| {{m_s} =  - 1} \right\rangle$ are encoded as a qubit which is manipulated by microwaves (MW).
		(c) Pulse sequence for qubit control and measurement. In the pulse sequence illustrated here, the microwave pulse in the measurement section aims for acquiring the return probability. The Bloch vector of the quantum state and the direction of the rotating field are shown below the pulse sequence.
	}
	\label{structure}
\end{figure}

%\emph{Experiment}.---
A negatively charged nitrogen-vacancy (NV) center in diamond is used in our experiment to simulate $H(k,t)$~\cite{GTPExp},
the many-body Hamiltonian in its quasimomentum representation.
As shown in Fig.~\ref{structure}(a), the NV center is composed of one substitutional nitrogen atom and an adjacent vacancy~\cite{Doherty,Schirhagl,Prawer,Wrachtrup}.
In our experiment, an external static magnetic field around $510$ G is parallel to the NV symmetry axis, which enables both the NV electron spin and the host $^{14}$N nuclear spin to be polarized by optical excitation~\cite{Jacques,Sar}.
The Hamiltonian of the electronic ground state of the NV center with a static magnetic field $B$ applied along the NV axis (also the $z$ axis) is $H_{\rm{NV}} = D S_z^2+\gamma B S_z$,
where $S_z$ is the angular momentum operator for spin-1, $D=2\pi\times2870$~MHz is the zero-field splitting, and $\gamma=2\pi\times2.8$~MHz/G is the gyromagnetic ratio of the NV electron.
As illustrated in Fig.~\ref{structure}(b), microwaves generated by an arbitrary waveform generator drives the transition between the electronic levels $\left| {{m_s} = 0} \right\rangle$ and $\left| {{m_s} =  - 1} \right\rangle$ which compose a qubit. The level $\left| {{m_s} =  1} \right\rangle$ remains idle due to large detuning.
%The Hamiltonian of the qubit in the lab frame is $H^{\rm{lab}} = \omega_0\sigma_{z}/2 + f(t)\sigma_{x}$,
%where the term $f(t)\sigma_{x}$ delineates the effect of the microwave field.
The probability in $\left| {{m_s} = 0} \right\rangle$ can be read out via fluorescence detection during optical excitation.
All the optical procedures are performed on a home-built confocal microscope, and a solid immersion lens is etched on the diamond above the NV center to enhance the fluorescence collection~\cite{Robledo,Rong}.
%In the following, the rotating frame determined by the resonant MW is adopted.

In our experiment, the evolution at different $k$ is performed separately in different experimental runs. The pulse sequence for each $k$ is sketched in Fig.~\ref{structure}(c).
At first, the qubit is polarized to the state $|0\rangle$ by a laser pulse [green bar in the preparation section in Fig.~\ref{structure}(c)].
A resonant microwave pulse [orange bar in Fig.~\ref{structure}(c)] is then applied to prepared the initial state as $|\varphi_{-}(k,0)\rangle$, which occupies the lower quasienergy band. The Bloch vector of this initial state is along the direction $\bm{n}(k,0)=\left( -\Omega \sin k,0,\omega- {\delta _1}\cos k  - {\delta _2}\right)$.
%The initial state is then prepared to be $|\varphi_{-}(k,0)\rangle$, i.e., at the lower quasienergy band.
%The Bloch vector of this initial state is along the direction $\bm{n}(k,0)=\left( -\Omega \sin k,0,\omega- {\delta _1}\cos k  - {\delta _2}\right)$, obtained by applying a resonance microwave pulse [orange bar in Fig.~\ref{structure}(c)] to the state $|0\rangle$.
The parameters adopted in our experiment are $\Omega = 2\pi \times 10$ MHz, $\omega = \delta_1 = 2\pi \times 5$ MHz, and $\delta_2= \pm 2\pi \times 5$ MHz. For $\delta_2=  2\pi \times 5$ MHz, the prepared initial state lies in the $x$-$y$ plane for $k_c=\pi/2$. This is not the case for any $k$ if $\delta_2= - 2\pi \times 5$ MHz, thus excluding DQPTs.

Upon initial-state preparation, the qubit is left to evolve under $H(k,t)$, namely, in the presence of a field with the transverse component $\Omega\sin k$ and the longitudinal component ${\delta _1}\cos k + {\delta _2}$. The field is rotated around the $z$ axis with the angular frequency $\omega$, which is implemented by applying a microwave pulse starting from $t=0$, as depicted by the blue bar in Fig.~\ref{structure}(c).

%and $\delta_2 = 2\pi \times 5$ or $-2\pi \times 5$ MHz.

\begin{figure}\centering
	\includegraphics[width=0.95\columnwidth]{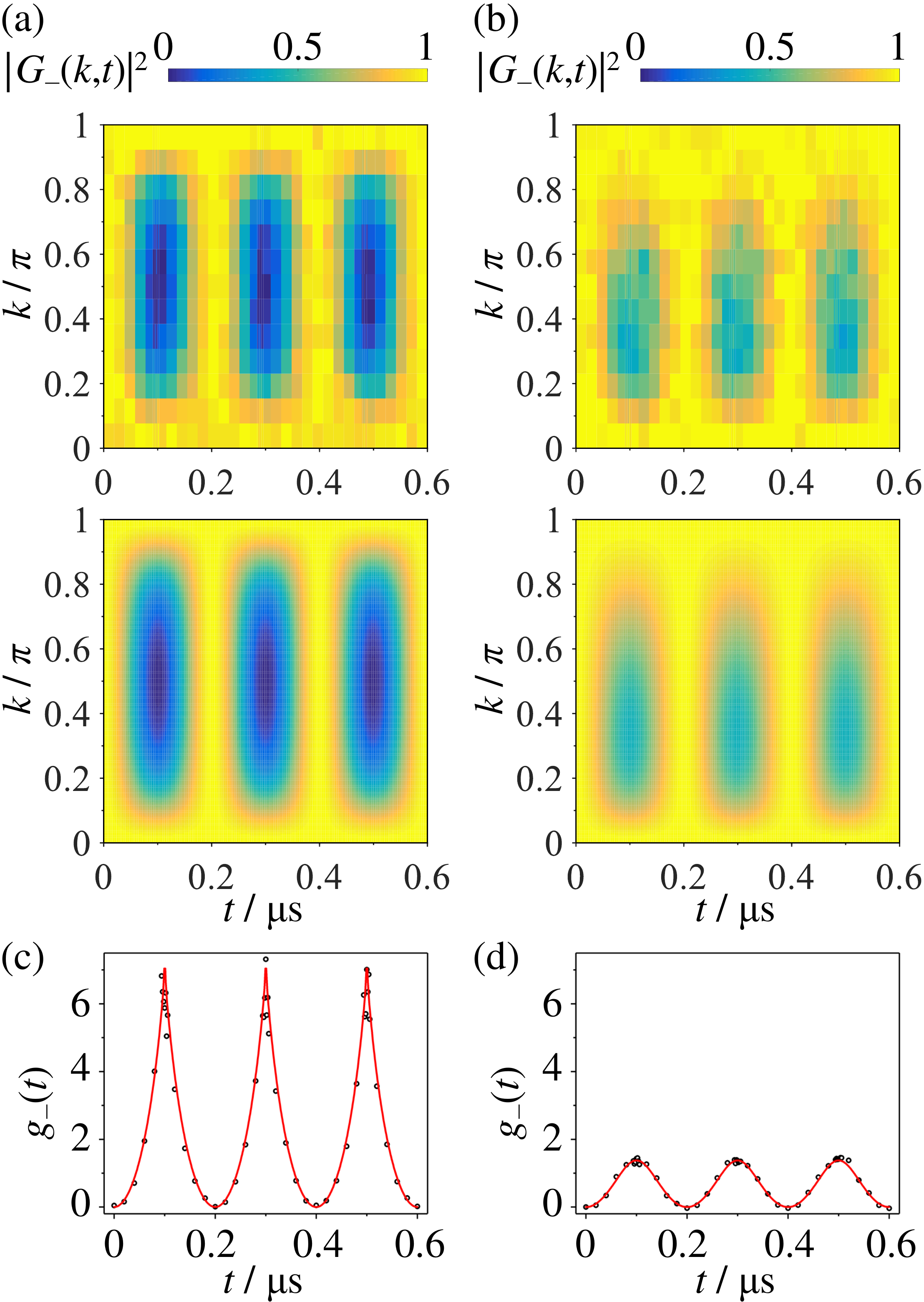}
	\caption{Return probabilities and rate functions.
		(a)(b) Return probabilities for $\delta_2=2\pi \times 5$ MHz and $-2\pi \times 5$ MHz, respectively.
		Experimental data are shown in the upper panels and theoretical results are presented in the lower panels.
		%Calculations based on the Schr\"{o}dinger equation are on the left panels, and experimental data are on the right panels.
		All panels in (a) and (b) share the same color bar.
		The associated rate functions are shown in panels (c) and (d),
		with black circles and red curves representing experimental data and theoretical values.
	}
	\label{rate}
\end{figure}

The evolution governed by $H(k,t)$ lasts for some duration $t_e$, and then the return probability $|G_-(k,t_e)|^2=|\langle \psi_-(k,0) | \psi_-(k,t_e) \rangle|^2$ needs to be measured.
The measurement procedure begins with a resonant microwave pulse [the magenta bar in Fig.~\ref{structure}(c)],
%comprises two steps, namely, a microwave pulse symbolized by the blue rectangle, and the subsequent laser illumination symbolized by the green rectangle together with fluorescence detection.
%The microwave pulse and the laser pulse are symbolized by the blue bar and the green rectangle, respectively, in the measurement section in Fig.~\ref{structure}(c).
%This pulse is described by $f(t)=\omega_1 \cos ( {\omega _0}t + \varphi_{\rm {I}} + \varphi_{\rm{II}}+ \varphi_{\rm{fin}})$, with $t$ starting from zero, $\varphi_{\rm{II}}=({\omega _0} - {\delta _1}\cos k  - {\delta _2})t_{\rm e}$, and $\varphi_{\rm{fin}} = \pi/2$. The duration of the pulse is $t_{\rm{fin}}=\alpha/\omega_1$.
which steers the direction of $\bm{n}(k,0)$ to the $+z$ direction. It is followed by a laser pulse [green bar in the measurement section of Fig.~\ref{structure}(c)] together with fluorescence detection.
%illumination together with fluorescence detection amounts to the measurement of $\sigma_z$.
The later is collected via two counting windows represented by the red bars in Fig.~\ref{structure}(c), with the first recording the signal while the second recording the reference~\cite{SM}.
The fluorescence collection amounts to the measurement of the probability in $|0\rangle$, and this effect combined with the last microwave pulse is equivalent to the measurement of $|G_-(k,t_{\rm e})|^2$.

The above sequence is performed for a series of $t_e$ within 0.6 $\mu$s, and is iterated five hundred thousand times to obtain the expectation value.
One can then get $|G_-(k,t)|^2$ as a function of $t$.
This procedure is repeated for different values of $k \in [0,\pi]$.
The experimental data with $\delta_2=2\pi \times 5$ MHz and $-2\pi \times 5$ MHz are illustrated in the upper panels of Fig.~\ref{rate}(a) and Fig.~\ref{rate}(b), respectively.
The experimental results agree with the theoretical ones which are shown in the lower panels.
%The condition (\ref{eq:CriticalMom}) is satisfied in the case of Fig.~\ref{rate}(a) but not satisfied in the case of Fig.~\ref{rate}(b).
The results in Fig.~\ref{rate}(a) confirm that, in the case of $\delta_2=2\pi \times 5$ MHz where the DQPT condition in Eq.~(\ref{eq:CriticalMom}) is satisfied, the return probability at the critical momentum $k_c=\pi/2$ vanishes at the critical times such as $t=0.1\ \rm{\mu s}$, $0.3\ \rm{\mu s}$, and $0.5\ \rm{\mu s}$.
The DQPT condition in Eq.~(\ref{eq:CriticalMom}) is not satisfied in the case of $\delta_2=-2\pi \times 5$ MHz.
The return probability never vanishes in this case as confirmed by the results in Fig.~\ref{rate}(b).

Numerical integration over $k$ based on the negative logarithm of these experimental data yields the experimental values of the rate function $g_-(t)$.
Under the condition in Eq.~(\ref{eq:CriticalMom}) for Floquet DQPTs, the rate function behaves non-analytically at critical times, as shown by the kinks in Fig.~\ref{rate}(c)~\cite{noerrorbar}. By contrast, the rate function $g_-(t)$ stretches smoothly over time without non-analyticity when the DQPT condition is not satisfied,
as shown in Fig.~\ref{rate}(d).
This further demonstrates that Floquet DQPTs can be observed through the rate function.

\begin{figure}\centering
	\includegraphics[width=1\columnwidth]{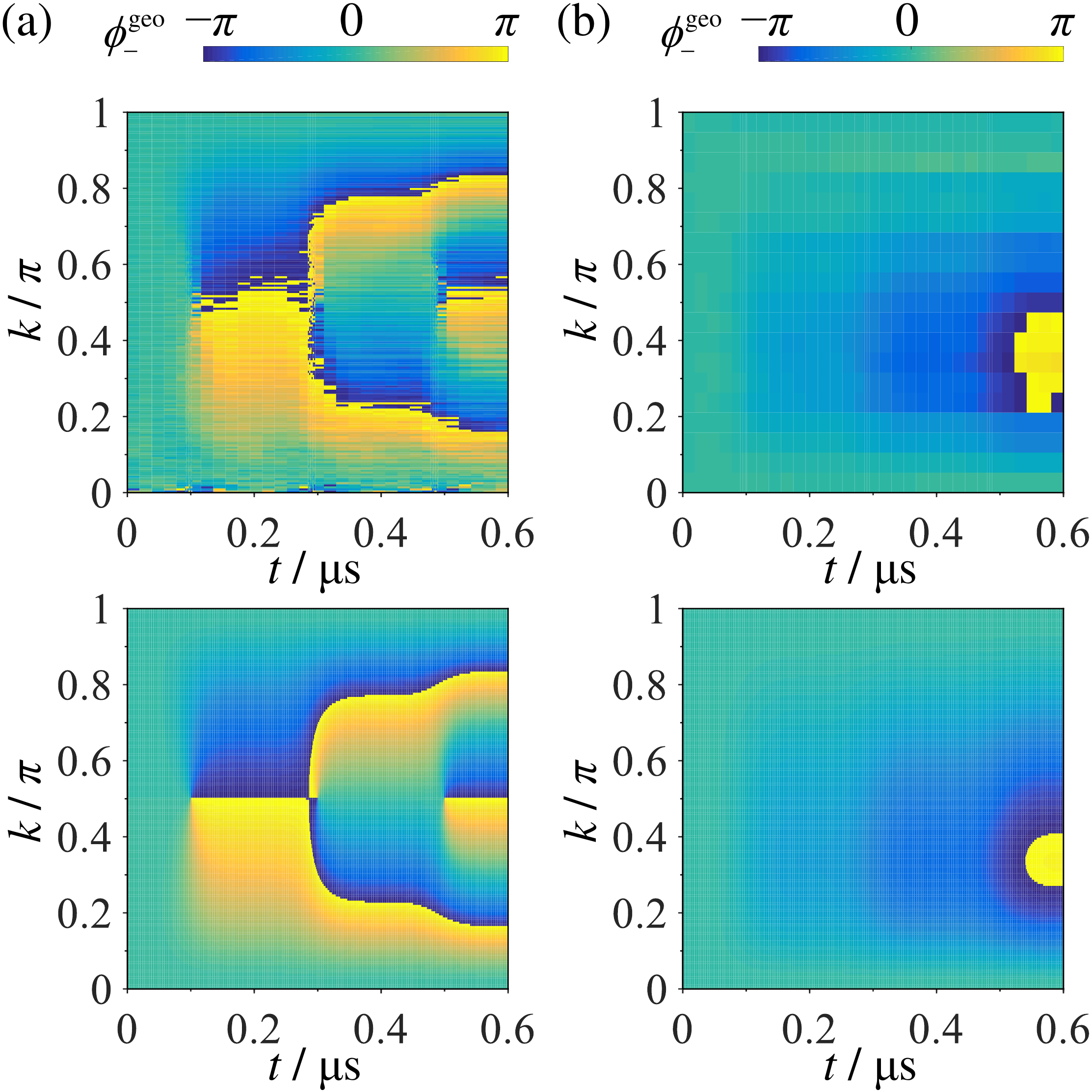}
	\caption{Measured geometric phases versus system parameter $k$ and time $t$ for (a) $\delta_2=2\pi \times 5$ MHz and (b) $\delta_2=-2\pi \times 5$ MHz. The bottom two panels represent theoretical results for the sake of comparison. In left panels, a change in the net jumps of the geometric phase along the $k$ dimension depicts a change in a dynamical topological invariant, thus signifying DQPTs at  $t=0.1\ \rm{\mu s}$,
 $0.3\ \rm{\mu s}$, and $ 0.5\ \rm{\mu s}$.  No such behavior is found in right panels.
	}
	\label{geometry}
\end{figure}

In order to investigate the topological character of Floquet DQPTs, we next extract the Pancharatnam geometric phase of the time-evolving state from the expectation values of $\sigma_x$, $\sigma_y$, and $\sigma_z$ measured in our experiment~\cite{SM}. The pulse sequence for measuring these expectation values is similar to that in Fig.~\ref{structure}(c) except for the last microwave pulse.
This resonant microwave pulse rotate the direction of $+x$ or $+y$ to $+z$ for the measurement of $\sigma_x$ or $\sigma_y$, respectively.
The pulse is not needed for the measurement of $\sigma_z$.
%The pulse for measurement realizes a $\pi/2$ rotation around the $-y$ ($+x$) direction to measure the expectation values of $\sigma_x$ ($\sigma_y$). The phase component $\varphi_{\rm{fin}}$ of this pulse is $-\pi/2$ (0) and the duration is $\pi/(2\omega_1)$. This pulse is absent while measuring the expectation values of $\sigma_z$.
The experimental and theoretical values of the geometric phases for $\delta_2=2\pi \times 5$ MHz and $-2\pi \times 5$ MHz are illustrated in Fig.~\ref{geometry}(a) and (b).
Geometric phases with $2\pi$ difference are equivalent and we have thus set them in the range from $-\pi$ to $\pi$.
%The dynamical topological invariant $\nu_-(t)$ in Eq.~\ref{eq:topology} arises from discontinuity points of the geometric phase in the $k$ dimension~\cite{SM}. The results are shown in Fig.~\ref{winding}(c).
The net discontinuous jump of the geometric phase along the $k$ dimension is a signature of the winding of the geometric phase with quasimomentum. The number of winding then yields $\nu _-(t)$ defined in Eq.~(\ref{eq:winding}) as a dynamical topological order parameter.
In the case of Fig.~\ref{geometry}(a) with Floquet DQPTs, the measured geometric phases manifest a jump from $-\pi$ to $\pi$ at the first critical time $t=0.1\ \rm{\mu s}$.
The number of net jumps increases by one once a new critical time is passed, in particular at $t=0.3\ \rm{\mu s}$ and $t=0.5\ \rm{\mu s}$.
Therefore, $\nu_-(t)$ increases from one at $t=0.1\ \rm{\mu s}$ to three at $t=0.5\ \rm{\mu s}$.
By contrast, in the case of Fig.~\ref{geometry}(b) where the DQPT condition in Eq.~(\ref{eq:CriticalMom}) is not satisfied and hence there are no DQPTs, the measured geometric phases has no net jumps along the $k$ dimension. Even when there is a discontinuous jump from $-\pi$ to $\pi$, it is cancelled by a reversed jump from $\pi$ to $-\pi$, resulting in no change in $\nu_-(t)$.
Some noisy but insignificant patterns in the experimental results are mainly due to the imperfection of the microwave pulses.
Nevertheless, even in the presence of such experimental errors, all the principal characteristics of Floquet DQPTs in connection with jumps in the dynamical topological order parameter have been clearly demonstrated.
%Additionally, in our model, the rate funtions with DQPT are all congruent under fixed $\Omega$, $\omega$, and $\delta1$.
%This character have also been supported by the experimental results.

%\emph{Summary}.---
In conclusion, we have discovered and experimentally demonstrated a new class of DQPTs as intrinsic features of quantum systems subject to smooth and periodic time modulations.   Floquet DQPTs here are found to have two-fold topological nature.
Firstly, their occurrence leads to jumps in a dynamical topological invariant.
Secondly, the system parameters yielding Floquet DQPTs lie in a regime accommodating topologically nontrivial Floquet phases.
Our work thus opens a door for future studies of DQPTs in connection with topological matter.

\begin{acknowledgements}
The authors at University of Science and Technology of China are supported by the National Key R$\&$D Program of China (Grant No.~2018YFA0306600, No.~2016YFA0502400),
the National Natural Science Foundation of China (Grants No.~81788101, No.~11227901, No.~31470835, No.~91636217, and No.~11722544),
the CAS (Grants No.~GJJSTD20170001, No.~QYZDY-SSW-SLH004, and No.~YIPA2015370),
Anhui Initiative in Quantum Information Technologies (Grant No.~AHY050000), the CEBioM, and the Fundamental Research Funds for the Central Universities (WK2340000064).
J.G. acknowledges support from the Singapore NRF Grant No.~NRF-NRFI2017-04 (WBS No.~R-144-000-378-281) and by the Singapore Ministry of Education Academic Research Fund Tier I (WBS No.~R-144-000-353-112).
L.Z. acknowledges support from the Young Talents Project at Ocean University of China (Grant No.~861801013196).

K.Y., L.Z., W.M., and X.K. contributed equally to this work.
\end{acknowledgements}

\onecolumngrid
\vspace{1.5cm}

\begin{center}
\textbf{\large Supplemental Material}
\end{center}

\setcounter{figure}{0}
\setcounter{equation}{0}
\makeatletter
\renewcommand{\thefigure}{S\@arabic\c@figure}
\renewcommand{\theequation}{S\@arabic\c@equation}
\renewcommand{\bibnumfmt}[1]{[S#1]}
\renewcommand{\citenumfont}[1]{S#1}

%\begin{abstract}
%~\\
%\tableofcontents
%\end{abstract}
%\maketitle

\section{I. Theory}
In this supplementary note, we present in detail our theory of Floquet dynamical quantum phase transitions~(DQPTs). We first briefly review the basics of conventional DQPTs and Floquet theory. Following that, we introduce our definition of Floquet DQPTs and discuss its general features. To demonstrate these features, we study the periodically driven spin chain model whose Hamiltonian in thermodynamic limit can be mapped to a qubit in a rotating field. We further solve this driven qubit model analytically and describe its Floquet DQPTs by (i) the Fisher zeros of the Loschmidt (return) amplitude, (ii) the rate function of Loschmidt echo (return probability), (iii) the pattern of a non-adiabatic, non-cyclic geometric phase, and (iv) a dynamical topological winding number. We also reveal the relation between Floquet DQPTs found in our model and its underlying Floquet topological phases. Finally, we illustrate numerically our theoretical predictions in three examples.

\subsection{A. Elements of DQPTs}\label{Sec:DQPTDef}
In this section, we briefly review the definition of DQPTs for quenched evolutions. According to Ref.~\cite{HeylPRL2013}, a DQPT means that the time derivative of certain observable at a given order shows a jump or a singularity in time. It is therefore a real-time non-analytic signature in dynamics. The central quantity in the description of DQPTs is the return amplitude, defined as
\begin{equation}
{\cal G}(t)=\langle\Psi_{i}|e^{-it {\cal H}_{f}}|\Psi_{i}\rangle,
\end{equation}
where $|\Psi_{i}\rangle$ is the initial state of the system, usually chosen to be the ground state of some many-body Hamiltonian ${\cal H}_{i}$, and ${\cal H}_{f}$ is the Hamiltonian governing the evolution of the system after the initial time $t=0$. A nonequilibrium dynamical process is generated if $[{\cal H}_{i},{\cal H}_{f}]\neq0$, which can be realized by performing a quantum quench from ${\cal H}_{i}$ to ${\cal H}_{f}$ at $t=0$.

Formally, ${\cal G}(t)$ mimics the so-called boundary partition function in equilibrium statistical mechanics, which can be seen by rotating the time $t$ to the complex plane ($it\rightarrow z=\tau+it\in\mathbb{C}$), yielding ${\cal G}(z)=\langle\Psi_{i}|e^{-z{\cal H}_{f}}|\Psi_{i}\rangle$~\cite{DQPTRev1}. The zeros of ${\cal G}(z)$ are called Fisher zeros. They form a dense set in thermodynamic limit. A real-time Fisher zero of ${\cal G}(z)$ appears when this set crosses the imaginary time axis at some critical time $t=t_{c}$, where the rate function of return probability,
\begin{equation}
g(t)\equiv -\lim_{N\rightarrow\infty}\frac{1}{N}\ln|{\cal G}(t)|^{2},
\end{equation}
or its time derivative becomes non-analytic as a function of time. Here $N$ is the number of degrees of freedom of the system. Mathematically, this is closely related to the microscopic Lee-Yang theory of phase transitions. Therefore ${\cal G}(t)$ and $g(t)$ are sometimes called dynamical partition function and dynamical free energy, respectively, whereas $t_{c}$ is called the critical time of a DQPT, analogous to the critical temperature of a thermal phase transition. Theoretical studies have shown that DQPTs usually happen in dynamics following a quench across the equilibrium quantum phase transition point of the prequench Hamiltonian~\cite{DQPTRev1}. Experimentally, DQPTs characterized by kinks in return rates at critical times were first observed in a one-dimensional chain of $10$ trapped ions~\cite{DQPTExp1}.

\subsection{B. Elements of Floquet theory}
In this section, we recap some basics of Floquet theory~\cite{HolthausFloTut} that will be used in this study. A Floquet system is described by a time periodic Hamiltonian
${\cal H}(t)={\cal H}(t+T)$, where $T$ is the driving period and $\omega=2\pi/T$
is the driving frequency. A complete set of solutions of the time-dependent Schr\"odinger equation is given by the Floquet states
$\{|\Psi_{n}(t)\rangle\}$, with each of them satisfying (take $\hbar=1$)
\begin{equation}
i\frac{d}{dt}|\Psi_{n}(t)\rangle={\cal H}(t)|\Psi_{n}(t)\rangle.
\end{equation}
The Floquet state $|\Psi_{n}(t)\rangle$ can be further decomposed as $|\Psi_{n}(t)\rangle=e^{-i\varepsilon_{n}t}|\Phi_{n}(t)\rangle$,
where $\varepsilon_{n}\in[-\omega/2,\omega/2)$ is called the quasienergy
and $|\Phi_{n}(t)\rangle$ is called the Floquet mode. A Floquet mode is time periodic, i.e.,
$|\Phi_{n}(t)\rangle=|\Phi_{n}(t+T)\rangle$.
All such Floquet modes form an orthonormal, complete set at any time
$t$, i.e.,
\begin{equation}
\sum_{\ell}|\Phi_{n}(t)\rangle\langle\Phi_{n}(t)|=1,\qquad\langle\Phi_{\ell}(t)|\Phi_{n}(t)\rangle=\delta_{\ell n}.
\end{equation}
In terms of the quasienergies and Floquet modes, the time evolution operator of the system from an initial time $0$ to a later time $t$ can be expressed as
\begin{equation}
{\cal U}(t)=\sum_{\ell}e^{-i\varepsilon_{\ell}t}|\Phi_{\ell}(t)\rangle\langle\Phi_{\ell}(0)|.
\label{FloquetOp}
\end{equation}
It is not hard to verify that $U(0)=1$ and
\begin{equation}
i\frac{d}{dt}{\cal U}(t)=\sum_{\ell}e^{-i\varepsilon_{\ell}t}\left(\varepsilon_{\ell}+i\frac{d}{dt}\right)|\Phi_{\ell}(t)\rangle\langle\Phi_{\ell}(0)|
=\sum_{\ell}e^{-i\varepsilon_{\ell}t}H(t)|\Phi_{\ell}(t)\rangle\langle\Phi_{\ell}(0)|=H(t){\cal U}(t),
\end{equation}
as expected. One may further write a Floquet mode $|\Phi_{n}(t)\rangle$
as $|\Phi_{n}(t)\rangle=P(t)|\Phi_{n}(0)\rangle$,
where $P(t)=P(t+T)$ is called the micromotion operator.

\subsection{C. Floquet DQPTs --- definition and properties}\label{FloquetDQPTTheory}
We will introduce our definition of Floquet DQPTs in this section.
Consider a system described by a time-periodic Hamiltonian ${\cal H}(t)$,
which is prepared initially ($t=0$) at an $N$-particle Floquet state $|\Psi_{n}(0)\rangle$, given by Floquet modes $|\Phi_n(0)\rangle$ filling the quasienergy band $n$.
At a later time $t$, the return amplitude of the system
to its initial state is given by
\begin{equation}
{\cal G}_n(t)=\langle\Psi_{n}(0)|{\cal U}(t)|\Psi_{n}(0)\rangle.
\end{equation}
The time evolution operator ${\cal U}(t)={\cal T}e^{-i\int_{0}^{t}ds {\cal H}(s)}$,
where ${\cal T}$ denotes the time ordering operator. In terms of quasienergies
and Floquet modes, ${\cal U}(t)$ can be expressed as Eq.~(\ref{FloquetOp}). Then we can
write the return amplitude for each Floquet mode involved in the dynamcis as
\begin{equation}
{\cal G}_{n}(t)=e^{-i\varepsilon_{n}t}\langle\Phi_{n}(0)|\Phi_{n}(t)\rangle.
\end{equation}
The corresponding return probability to the initial state is given by
\begin{equation}
|{\cal G}_{n}(t)|^{2}=|\langle\Phi_{n}(0)|\Phi_{n}(t)\rangle|^{2}=|\langle\Phi_{n}(0)|P(t)|\Phi_{n}(0)\rangle|^{2},
\label{GAbsSquare}
\end{equation}
with the micromotion operator $P(t)$ introduced in the previous section. In thermodynamic
limit, the rate function of return probability reads
\begin{equation}
g_{n}(t)=-\lim_{N\rightarrow\infty}\frac{1}{N}\ln|{\cal G}_n(t)|^{2}=-\lim_{N\rightarrow\infty}\frac{1}{N}\sum_{\rm{occ.}}\ln|{\cal G}_n(t)|^{2},
\label{gnt}
\end{equation}
where $N$ is the number of degrees of freedom (e.g., number of particles) of the system and the summation $\sum_{\rm{occ.}}$~is taken over the $N$ occupied states. In our construction of Floquet DQPTs,
$g_{n}(t)$ plays the role of a dynamical free energy. If there exists
a critical time $t_{c}$, at which $|\Phi_{n}(t_{c})\rangle$
is orthogonal to the initial state $|\Phi_{n}(0)\rangle$, ${\cal G}(t_{c})$ will vanishes.
Then $g_{n}(t)$ as defined in Eq.~(\ref{gnt}) or its time derivatives will become non-analytic as a function of time $t$.
We refer to this as a Floquet DQPT. From Eq.~(\ref{GAbsSquare}),
one may interpret a Floquet DQPT as originated from the interplay
between the stroboscopic nature of the system encoded in its Floquet
mode $|\Phi_{n}(0)\rangle$, and the micromotion within a driving period described by $P(t)$.

Compared with the DQPTs following a sudden quench~\cite{HeylPRL2013}, a notable
difference of Floquet DQPTs is that they happen in time periodically. This is not hard to see from Eq.~(\ref{gnt}), since
\begin{alignat}{1}
g_{n}(t+T)= & -\lim_{N\rightarrow\infty}\frac{1}{N}\ln|{\cal G}_n(t+T)|^{2}
=-\lim_{N\rightarrow\infty}\frac{1}{N}\sum_{\rm{occ.}}\ln|\langle\Phi_{n}(0)|P(t+T)|\Phi_{n}(0)\rangle|^{2}\nonumber \\
= & -\lim_{N\rightarrow\infty}\frac{1}{N}\sum_{\rm{occ.}}\ln|\langle\Phi_{n}(0)|P(t)|\Phi_{n}(0)\rangle|^{2}=g_{n}(t).
\end{alignat}
Therefore, if $t_{c}$ is a critical time, then $t_{c}+\nu T$ is also
a critical time for any $\nu\in\mathbb{Z}$, with $g_{n}(t)$ taking
equal values at these critical times. Practically, this means that
Floquet DQPTs will not just happen in a transient time scale as the
DQPTs following a single quench, but will be observable with equal
strength in a much wider time window thanks to the periodic
drivings applied to the system. Furthermore, the driving fields also
yield more freedom to control Floquet DQPTs in the system.

In the following sections of this theoretical part, we will study an
analytically solvable driven spin chain model as described in the main text, which possesses the
defining features of Floquet DQPTs. We will further map the Hamiltonian
of this driven spin chain to the parameter space of a qubit in a rotating
magnetic field, which allows us to develop a systematic description of Floquet
DQPTs in this system (see Table~\ref{Tab1} for an outline).

\begin{table}
	\begin{centering}
		\begin{tabular}{|ccc|}
			\hline
			\textbf{Concepts} & \textbf{Expressions} & \textbf{Annotations}\tabularnewline
			\hline
			\hline
			Spin chain  & ${\cal H}(t)=\sum_{n}\left\{ [\delta_{1}-\Omega\sin(\omega t)]s_{n}^{x}s_{n+1}^{x}+[\delta_{1}+\Omega\sin(\omega t)]s_{n}^{y}s_{n+1}^{y}\right\} $ & \multicolumn{1}{c|}{${\cal H}(t)={\cal H}(t+T)$}\tabularnewline
			Hamiltonian & $-\delta_{2}\sum_{n}s_{n}^{z}-\Omega\cos(\omega t)\sum_{n}\left(s_{n}^{x}s_{n+1}^{y}+s_{n}^{y}s_{n+1}^{x}\right)$ & $T=2\pi/\omega$\tabularnewline
			\hline
			Lattice & ${\cal H}(t)=\sum_{k\in{\rm BZ}}\Psi_{k}^{\dagger}H(k,t)\Psi_{k}$ & Obtained from the spin chain Hamiltonian through\tabularnewline
			Hamiltonian & $H(k,t)=\Omega\sin k[\cos(\omega t)\sigma_{x}+\sin(\omega t)\sigma_{y}]/2+(\delta_{1}\cos k + \delta_{2})\sigma_{z}/2$ & Jordan-Wigner and Fourier transforms,\tabularnewline
			%&  & $H(k,t)$ is realized experimentally\tabularnewline
			\hline
			Bloch Hamiltonian & \multirow{2}{*}{$H_{F}(k)\equiv U_{R}^{\dagger}(t)\left[H(k,t)-i\frac{d}{dt}\right]U_{R}(t)$} & \multirow{2}{*}{Rotation $U_{R}(t)=\begin{pmatrix}1 & 0\\
				0 & e^{i\omega t}
				\end{pmatrix}=U_{R}(t+T)$}\tabularnewline
			in rotating frame &  & \tabularnewline
			\hline
			Quasienergies and & $E_{\pm}(k)={\omega}/{2}\pm\sqrt{h_{xy}^{2}(k)+\left[h_{z}(k)-{\omega}/{2}\right]^{2}}$ & $E_{\pm}(k)$ and $|\varphi_{\pm}(k,0)\rangle$ are eigenvalues\tabularnewline
			Floquet eigenstates & $|\psi_{\pm}(k,t)\rangle=e^{-iE_{\pm}(k)t}U_{R}(t)|\varphi_{\pm}(k,0)\rangle$ & and eigenstates of $H_{F}(k)$\tabularnewline
			\hline
			Initial & \multirow{2}{*}{$|\Psi_{\alpha}(0)\rangle=\prod_{k>0}|\psi_{\alpha}(k,0)\rangle$} & \multirow{2}{*}{Fill the Floquet band $\alpha=+$ or $-$}\tabularnewline
            state &  & \tabularnewline
			\hline
			\textcolor{cyan}{Return} & \multirow{2}{*}{$G_{\alpha}(k,t)\equiv\langle\psi_{\alpha}(k,0)|\psi_{\alpha}(k,t)\rangle$} & \multirow{2}{*}{``Dynamical partition function''}\tabularnewline
			\textcolor{cyan}{amplitude} &  & \tabularnewline
			\hline
			\textcolor{cyan}{Rate} & \multirow{2}{*}{$g_{\alpha}(t)=-\int_{0}^{\pi} dk\ln|{\cal G}_{\alpha}(k,t)|^{2}/\pi=g_{\alpha}(t+T)$} & \multirow{2}{*}{``Dynamical free energy''}\tabularnewline
			\textcolor{cyan}{function} &  & \tabularnewline
			\hline
			\textcolor{cyan}{Critical time and} & $t_{c}=(2n-1)T/2,\qquad n\in\mathbb{Z}^{+}$ & Floquet DQPTs happen at each $t_{c}$\tabularnewline
			\textcolor{cyan}{critical momentum} & $k_{c}=\arccos\left[\left(\omega-\delta_{2}\right)/\delta_{1}\right]$ & if there exists such a $k_{c}\in[0,\pi]$\tabularnewline
			\hline
			\textcolor{cyan}{Pancharatnam } & \multirow{2}{*}{$\phi_{\alpha}^{{\rm geo}}(k,t)=-i\ln\left[\frac{G_{\alpha}(k,t)}{|G_{\alpha}(k,t)|}\right]+\langle\varphi_{\alpha}(k)|H_{R}(k)|\varphi_{\alpha}(k)\rangle t$} & \multirow{2}{*}{Total phase minus dynamical phase}\tabularnewline
			\textcolor{cyan}{geometric phase} &  & \tabularnewline
			\hline
			\textcolor{cyan}{Dynamical} & \multirow{2}{*}{$\nu_{\alpha}(t)=\int_{0}^{\pi}{dk}\left[\partial_{k}\phi_{\alpha}^{{\rm geo}}(k,t)\right]/({2\pi})$} & Topological order parameter \tabularnewline
			\textcolor{cyan}{winding number} &  & of the Floquet DQPTs\tabularnewline
			\hline
		\end{tabular}
		\par\end{centering}
	\caption{Outline for the concepts introduced in Sections I.D and I.E. Key quantities characterizing Floquet DQPTs are highlighted in cyan.}
	\label{Tab1}
\end{table}

\subsection{D. Floquet DQPTs in a periodically driven spin chain}
We start with a periodically driven spin chain with the Hamiltonian
\begin{alignat}{1}
{\cal H}(t)&=  \sum_{n}\left\{ [\delta_{1}-\Omega\sin(\omega t)]s_{n}^{x}s_{n+1}^{x}+[\delta_{1}+\Omega\sin(\omega t)]s_{n}^{y}s_{n+1}^{y}\right\} -\delta_{2}\sum_{n}s_{n}^{z}\label{XYStatic}\\
&-  \Omega\cos(\omega t)\sum_{n}\left(s_{n}^{x}s_{n+1}^{y}+s_{n}^{y}s_{n+1}^{x}\right)
\label{XYDrive}
\end{alignat}
where $\delta_{1}$, $\delta_{2}$ and $\Omega$ are real parameters, $\omega=2\pi/T$ is the driving frequency with $T$ being the driving period, and $s_{n}^{x,y,z}$ are quantum spin-$1/2$
variables located at site $n$ of the spin chain.
The terms on the right hand side of Eq.~(\ref{XYStatic}) describes the usual XY model with time-dependent nearest-neighbor couplings,
whereas the terms in Eq.~(\ref{XYDrive}) are some anomalous coupling terms. In terms of Pauli spin variables $\sigma_{n}^{x,y,z}=2s_{n}^{x,y,z}$,
the spin chain Hamiltonian can be expressed as
\begin{alignat}{1}
{\cal H}(t)&=  \sum_{n}\left[\frac{\delta_{1}-\Omega\sin(\omega t)}{4}\sigma_{n}^{x}\sigma_{n+1}^{x}+\frac{\delta_{1}+\Omega\sin(\omega t)}{4}\sigma_{n}^{y}\sigma_{n+1}^{y}\right]-\frac{\delta_{2}}{2}\sum_{n}\sigma_{n}^{z}\\
&-  \frac{\Omega\cos(\omega t)}{4}\sum_{n}\left(\sigma_{n+1}^{x}\sigma_{n}^{y}+\sigma_{n}^{x}\sigma_{n+1}^{y}\right).
\end{alignat}
Pauli spins are anticommute at the same site, i.e. $\{\sigma_{n}^{\alpha},\sigma_{n}^{\beta}\}=2\delta_{\alpha\beta}$,
but commute at different sites, i.e. $[\sigma_{m}^{\alpha},\sigma_{n}^{\beta}]=2i\delta_{mn}\epsilon_{\alpha\beta\gamma}\sigma_{n}^{\gamma}$,
with $\alpha,\beta,\gamma\in x,y,z$.
In the following, we first fermionize ${\cal H}(t)$ by applying the Jordan-Wigner transformation, and then apply the Fourier transform to the resulting free fermionic lattice model to obtain a single qubit Hamiltonian in momentum representation under periodic boundary conditions.

\subsubsection{1. Fermionization of the spin chain}

The Jordan-Wigner transformation is a non-local transformation. It
is often used to exactly solve 1D spin chains such as the Ising and
XY models by transforming the spin operators to fermionic operators,
and then doing diagonalizations in the fermionic basis \cite{FranchiniBook}.

We first express $\sigma_{n}^{x}$ and $\sigma_{n}^{y}$ in terms
of spin raising and lowering operators $\sigma_{n}^{\pm}=(\sigma_{n}^{x}\pm i\sigma_{n}^{y})/2$ as
\begin{equation}
\sigma_{n}^{x}=\sigma_{n}^{+}+\sigma_{n}^{-},\qquad\sigma_{n}^{y}=-i(\sigma_{n}^{+}-\sigma_{n}^{-}).
\end{equation}
In terms of $\sigma_{n}^{\pm}$ and $\sigma_{n}^{z}$, the spin chain
Hamiltonian reads
\begin{equation}
{\cal H}(t)=  \frac{\delta_{1}}{2}\sum_{n}(\sigma_{n}^{+}\sigma_{n+1}^{-}+\sigma_{n}^{-}\sigma_{n+1}^{+})-\frac{\delta_{2}}{2}\sum_{n}\sigma_{n}^{z}+  \frac{\Omega}{2i}\sum_{n}\left(e^{-i\omega t}\sigma_{n}^{-}\sigma_{n+1}^{-}-e^{i\omega t}\sigma_{n}^{+}\sigma_{n+1}^{+}\right).
\end{equation}
The Jordan-Wigner transformation~(JWT) is defined as
\begin{equation}
\sigma_{n}^{+}=  e^{i\pi\sum_{m<n}f_{m}^{\dagger}f_{m}}f_{n},~~~~~~
\sigma_{n}^{-}=  f_{n}^{\dagger}e^{-i\pi\sum_{m<n}f_{m}^{\dagger}f_{m}},~~~~~~
\sigma_{n}^{z}=  1-2f_{n}^{\dagger}f_{n},
\end{equation}
where the fermionic creation and annihilation operators $f_{n}^{\dagger}$
and $f_{n}$ satisfy the anti-commutation relations
\begin{equation}
\{f_{m},f_{n}\}=\{f_{m}^{\dagger},f_{n}^{\dagger}\}=0,\qquad\{f_{m},f_{n}^{\dagger}\}=\delta_{mn}.
\end{equation}
Using the JWT, we find
\begin{equation}
\begin{aligned}
\sigma_{n}^{+}\sigma_{n+1}^{-}&=  e^{i\pi\sum_{m<n}f_{m}^{\dagger}f_{m}}f_{n}f_{n+1}^{\dagger}e^{-i\pi\sum_{m<n+1}f_{m}^{\dagger}f_{m}} =  f_{n}f_{n+1}^{\dagger}e^{i\pi\sum_{m<n}f_{m}^{\dagger}f_{m}}e^{-i\pi\sum_{m<n+1}f_{m}^{\dagger}f_{m}}\\
&=  -f_{n+1}^{\dagger}f_{n}e^{-i\pi f_{n}^{\dagger}f_{n}}=f_{n+1}^{\dagger}f_{n},
\end{aligned}
\end{equation}
\begin{equation}
\sigma_{n}^{-}\sigma_{n+1}^{+}=(\sigma_{n}^{+}\sigma_{n+1}^{-})^{\dagger}=f_{n}^{\dagger}f_{n+1},
\end{equation}
\begin{equation}
\begin{aligned}
\sigma_{n}^{-}\sigma_{n+1}^{-}&=  f_{n}^{\dagger}e^{-i\pi\sum_{m<n}f_{m}^{\dagger}f_{m}}f_{n+1}^{\dagger}e^{-i\pi\sum_{m<n+1}f_{m}^{\dagger}f_{m}}
=  f_{n}^{\dagger}f_{n+1}^{\dagger}e^{-i\pi\sum_{m<n}f_{m}^{\dagger}f_{m}}e^{-i\pi\sum_{m<n+1}f_{m}^{\dagger}f_{m}} \\
&=  -f_{n+1}^{\dagger}f_{n}^{\dagger}e^{-i\pi f_{n}^{\dagger}f_{n}}=f_{n}^{\dagger}f_{n+1}^{\dagger},
\end{aligned}
\end{equation}
and
\begin{equation}
\sigma_{n}^{+}\sigma_{n+1}^{+}=(\sigma_{n}^{-}\sigma_{n+1}^{-})^{\dagger}=f_{n+1}f_{n}.
\end{equation}
Using these relations, we find the fermionized spin chain model to be
\begin{equation}
{\cal H}(t)=\frac{\delta_{1}}{2}\sum_{n}(f_{n}^{\dagger}f_{n+1}+{\rm h.c.})+\frac{\delta_{2}}{2}\sum_{n}(2f_{n}^{\dagger}f_{n}-1)+\frac{\Omega}{2i}\sum_{n}\left(e^{-i\omega t}f_{n}^{\dagger}f_{n+1}^{\dagger}-{\rm h.c.}\right).
\end{equation}
Dropping the constant term $({\delta_{2}}/{2})\sum_{n}1$, which only introduces a global shift to the energy, we arrive at the following quadratic fermionic Hamiltonian
\begin{equation}
{\cal H}(t)=\frac{\delta_{1}}{2}\sum_{n}(f_{n}^{\dagger}f_{n+1}+{\rm h.c.})+\delta_{2}\sum_{n}f_{n}^{\dagger}f_{n}+\frac{\Omega}{2i}\sum_{n}\left(e^{-i\omega t}f_{n}^{\dagger}f_{n+1}^{\dagger}-{\rm h.c.}\right).
\label{FermiLattice}
\end{equation}
The first term on the RHS of Eq.~(\ref{FermiLattice}) describes a nearest neighbor
(NN) hopping of fermions with hopping amplitude ${\delta_{1}}/{2}$.
The second term describes an onsite potential with strength $\delta_{2}$.
The third term describes superconducting pairing interactions between
NN fermions, with a complex pairing strength ${\Omega}/({2i})$
modulated periodically in time with the period $T=2\pi/\omega$.

\subsubsection{2. Mapping a fermionic chain to a qubit}

Under the anti-periodic boundary condition $f_{n}=-f_{n+N}$~\cite{FranchiniBook}, the fermionic
Hamiltonian Eq.~(\ref{FermiLattice}) can be further simplified by performing a Fourier
transform, where $N$ is the total number of lattice sites.

The Fourier transform is given by
\begin{equation}
f_{n}=\frac{1}{\sqrt{N}}\sum_{k\in{\rm{BZ}}}e^{ikn}f_{k},\qquad f_{n}^{\dagger}=\frac{1}{\sqrt{N}}\sum_{k\in{\rm{BZ}}}e^{-ikn}f_{k}^{\dagger},
\end{equation}
where the quasimomentum $k$ is in the first Brillouin zone~(BZ), and the lattice constant has be set to $1$. Using this transform, we find
\begin{equation}
\sum_{n=1}^Nf_{n}^{\dagger}f_{n}=\sum_{k\in{\rm BZ}}f_{k}^{\dagger}f_{k},\qquad\sum_{n=1}^Nf_{n}^{\dagger}f_{n+1}=\sum_{k\in{\rm BZ}}f_{k}^{\dagger}f_{k}e^{ik},\qquad\sum_{n=1}^Nf_{n}^{\dagger}f_{n+1}^{\dagger}=\sum_{k\in{\rm BZ}}e^{ik}f_{k}^{\dagger}f_{-k}^{\dagger}.\label{FTofH}
\end{equation}
Plugging Eq.~(\ref{FTofH}) into Eq.~(\ref{FermiLattice}), we find
\begin{alignat}{1}
{\cal H}(t)&=  \sum_{k\in{\rm BZ}}[\delta_{1}\cos(k)+\delta_{2}]f_{k}^{\dagger}f_{k}+\frac{\Omega}{2i}\sum_{k}\left(e^{-i\omega t}e^{ik}f_{k}^{\dagger}f_{-k}^{\dagger}-e^{i\omega t}e^{-ik}f_{-k}f_{k}\right)\nonumber \\
&=  \sum_{k\in{\rm BZ}}\left\{ [\delta_{1}\cos(k)+\delta_{2}]f_{k}^{\dagger}f_{k}+\frac{\Omega\sin(k)}{2}\left(e^{-i\omega t}f_{k}^{\dagger}f_{-k}^{\dagger}+e^{i\omega t}f_{-k}f_{k}\right)\right\} .
\end{alignat}
In terms of the spinor basis $\Psi_{k}^{\dagger}=\begin{pmatrix}f_{k}^{\dagger} & f_{-k}\end{pmatrix}$,
we can further express ${\cal H}(t)$ as (up to a constant)
\begin{equation}
{\cal H}(t)=\sum_{k\in{\rm BZ}}\Psi_{k}^{\dagger}H(k,t)\Psi_{k},\label{Ht}
\end{equation}
where the Hamiltonian $H(k,t)$ is given by
\begin{equation}
H(k,t)=h_{xy}(k)[\cos(\omega t)\sigma_{x}+\sin(\omega t)\sigma_{y}]+h_{z}(k)\sigma_{z},
\label{eq:Hqubit}
\end{equation}
with
\begin{equation}
h_{xy}(k)=\frac{\Omega\sin k}{2},~~~~~~h_{z}(k)=\frac{\delta_{1}\cos k+\delta_{2}}{2}.
\end{equation}

Notably, the Hamiltonian $H(k,t)$ is exactly the Hamiltonian realized in the single-qubit simulation of generalized Thouless pump
\cite{GTPExp}, where the quasimomentum $k$ is mapped to angles of the qubit on the Bloch sphere. Furthermore, as will be shown in the next section,
the dynamics described by $H(k,t)$ is analytically solvable. Therefore this model serves as an ideal playground to explore Floquet DQPTs.

\subsection{E. Simulation of Floquet DQPTs by a driven qubit}

The Hamiltonian $H(k,t)$ in Eq.~(\ref{eq:Hqubit}) also describe a qubit in a rotating field.
%with field amplitude $h_{xy}(k)=\Omega\sin(k)$ in $x$-$y$ plane and $h_{z}(k)=\delta_{1}\cos(k)+\delta_{2}$ along $z$-axis.
%The parameter $k$ has the physical meaning of quasimomentum in the original fermionized problem, which is a conserved quantity.
The dynamics of the qubit obeys the Schr\"odinger equation
\begin{equation}
i\frac{d}{dt}|\psi(k,t)\rangle=H(k,t)|\psi(k,t)\rangle.
\label{Eq:evo}
\end{equation}
In the rotating frame defined by
\begin{equation}
U_{R}(t)=\begin{pmatrix}1 & 0\\
0 & e^{i\omega t}
\end{pmatrix},
\label{Eq:UR}
\end{equation}
the Hamiltonian in Eq.~(\ref{eq:Hqubit}) is transformed to
\begin{equation}
%H_{F}(k)\equiv U_{R}^{\dagger}(t)H(k,t)U_{R}(t)-iU_{R}^{\dagger}(t)\frac{d}{dt}U_{R}(t)=\frac{h_{xy}(k)}{2}\sigma_{x} + \frac{h_{z}(k)-\omega}{2}\sigma_{z}+\frac{\omega}{2}\sigma_{0},
H_{F}(k)\equiv U_{R}^{\dagger}(t)H(k,t)U_{R}(t)-iU_{R}^{\dagger}(t)\frac{d}{dt}U_{R}(t)=h_{xy}(k)\sigma_{x} + \left[h_{z}(k)-\frac{\omega}{2}\right]\sigma_{z}+\frac{\omega}{2}\sigma_{0},
\label{Eq:staticH}
\end{equation}
where $\sigma_{0}$ is a $2\times2$ identity matrix.
with the rotated state
\begin{equation}
|\varphi(k,t)\rangle=U_{R}^{\dagger}(t)|\psi(k,t)\rangle.
\end{equation}
The eigenvalues of the Hamiltonian in Eq.~(\ref{Eq:staticH}) are
\begin{equation}
E_{\pm}(k)=\frac{\omega}{2}\pm\sqrt{h_{xy}^{2}(k)+\left[h_{z}(k)-\frac{\omega}{2}\right]^{2}}.
\label{Eq:Epm}
\end{equation}
The eigenstates are
\begin{equation}
|\chi_{+}(k)\rangle=\frac{1}{\sqrt{2}}\begin{bmatrix}\sqrt{1+\frac{2h_{z}(k)-\omega}{\Delta(k)}}\\
{\rm sgn}(h_{xy})\sqrt{1-\frac{2h_{z}(k)-\omega}{\Delta(k)}}
\end{bmatrix},~~~~~~
|\chi_{-}(k)\rangle=\frac{1}{\sqrt{2}}\begin{bmatrix}{\rm sgn}(h_{xy})\sqrt{1-\frac{2h_{z}(k)-\omega}{\Delta(k)}}\\
-\sqrt{1+\frac{2h_{z}(k)-\omega}{\Delta(k)}}
\end{bmatrix},
\label{Eq:VarphiPM}
\end{equation}
with the energy gap $\Delta(k) = E_+(k)-E_-(k)=2\sqrt{h_{xy}^{2}(k)+\left[h_{z}(k)-\omega/2\right]^{2}}$.

The solution of the Schr\"odinger equation~(\ref{Eq:evo}) can be written as
\begin{equation}
|\psi(k,t)\rangle=U_{R}(t)e^{-iH_F(k)t}|\psi(k,0)\rangle.
\label{GenSol}
\end{equation}
The Floquet states and the Floquet modes are
\begin{equation}
|\psi_{\pm}(k,t)\rangle=e^{-iE_{\pm}(k)t}|\varphi_{\pm}(k,t)\rangle,~~~~~~|\varphi_{\pm}(k,t)\rangle=U_{R}(t)|\chi_{\pm}(k)\rangle.
\label{FloquetSol}
\end{equation}

With these results, we are ready to explore Floquet DQPTs through
this driven qubit model. Following our definitions in Sec.~\ref{FloquetDQPTTheory}, we choose the
initial state of the system to fill one of the Floquet band, i.e.,
\begin{equation}
|\Psi_{\pm}(0)\rangle=\prod_{k>0}|\psi_{\pm}(k,0)\rangle=\prod_{k>0}|\varphi_{\pm}(k,0)\rangle.
\end{equation}
Due to the conservation of $k$, the dynamics of Floquet states with
different $k$ can be treated separately. We will keep this in mind
and derive relevant quantities to characterize Floquet DQPTs in the
following subsections.

\subsubsection{1. Return amplitude}
For a state initialized at $t=0$ in a specific Floquet-Bloch state $|\psi_{\pm}(k,0)\rangle$,
its return amplitude at a later time $t$ is given by
\begin{equation}
G_{\pm}(k,t)\equiv\langle\psi_{\pm}(k,0)|\psi_{\pm}(k,t)\rangle=e^{-iE_{\pm}(k)t}\langle\varphi_{\pm}(k,0)|U_{R}(t)|\varphi_{\pm}(k,0)\rangle,
\label{RTAmp}
\end{equation}
where we have used Eq.~(\ref{FloquetSol}) to arrive at the second equality. According to our discussion in Sec.~\ref{FloquetDQPTTheory}, the condition for a Floquet
DQPT to happen is that there exist some quasimomentum $k$ and time
$t$ at which ${\cal G}_{\pm}(k,t)$ vanishes. Since the factor
$e^{-iE_{\pm}(k)t}$ can never be zero, this condition is equivalent
to $\langle\varphi_{\pm}(k,0)|U_{R}(t)|\varphi_{\pm}(k,0)\rangle=0$.
Using Eqs.~(\ref{Eq:UR}) and (\ref{Eq:VarphiPM}), this condition is explicitly given by
\begin{equation}
\frac{h_{xy}^{2}(k)+e^{i\omega t}[E_{\pm}(k)-h_{z}(k)]^{2}}{h_{xy}^{2}(k)+[E_{\pm}(k)-h_{z}(k)]^{2}}=0.
\label{Eq:FloquetDQPTCond}
\end{equation}
Physically, one may interpret this as requiring the average of micromotion
effects described by $U_{R}(t)$ to vanish, which will happen when
the initial state evolves to its orthogonal state.

\subsubsection{2. Fisher zeros}

The critical momenta and times of Floquet DQPTs are determined by
the Fisher zeros of the return amplitude $G_{\pm}(k,t)$. To
find these zeros on the complex plane, we extend $it\rightarrow z_{\pm}=\tau_{\pm}+it_{\pm}$,
where $\tau_{\pm}$ and $t_{\pm}$ are all real numbers. The condition Eq.~(\ref{Eq:FloquetDQPTCond}) can then be expressed as
\begin{equation}
e^{\omega\tau_{\pm}+i\omega t_{\pm}}=-\frac{h_{xy}^{2}(k)}{[E_{\pm}(k)-h_{z}(k)]^{2}},
\end{equation}
whose solutions $z_{\pm}$ are
\begin{equation}
\tau_{\pm}+it_{\pm}=\frac{1}{\omega}\ln\left\{ \frac{h_{xy}^{2}(k)}{[E_{\pm}(k)-h_{z}(k)]^{2}}\right\} +i(2n-1)\frac{\pi}{\omega}.
\end{equation}
Since the appearance of a Floquet DQPT is related to the existence of a Fisher zero on the imaginary axis ($\tau_\pm=0$), the critical momentum $k_{c}$ is given by the positive
$k$ solutions of
\begin{equation}
{\rm Re}z_{\pm}=\tau_{\pm}(k)=\frac{1}{\omega}\ln\left\{ \frac{h_{xy}^{2}(k)}{[E_{\pm}(k)-h_{z}(k)]^{2}}\right\} =0,
\label{Eq:CriticalK}
\end{equation}
and the critical times are determined by
\begin{equation}
{\rm Im}z_{\pm}=t_{\pm}=(2n-1)\frac{\pi}{\omega}=\frac{2n-1}{2}T,\qquad n\in\mathbb{Z}^{+}.
\end{equation}
As an important feature, the critical times in our model are independent
of $k$, which means that each line of Fisher zeros with a given $n$
is parallel to the real axis in the complex $z_{\pm}$ plane. Whether or not there exists
a Floquet DQPT is determined by whether a given line of Fisher zeros
could go across the imaginary axis, and therefore whether there
is a positive critical momentum $k=k_{c}$ which solves Eq.~(\ref{Eq:CriticalK}).

To summarize, our system prepared at a filled Floquet band ($|\Psi_{\pm}(0)\rangle=\prod_{k>0}|\psi_{\pm}(k,0)\rangle$) will
undergo Floquet DQPTs if there exist a $k=k_{c}$ such that $\tau_{\pm}(k_{c})=0$.
The corresponding Floquet DQPTs happen at critical times $t_{\pm}=(2n-1)T/2$
(i.e., at each midpoint of a driving period) for all $n\in\mathbb{Z}^{+}$.

\subsubsection{3. Explicit condition of Floquet DQPTs in the driven qubit model}
To have an explicit expression for the critical momentum $k_{c}$,
we need to find the solution of Eq.~(\ref{Eq:CriticalK}), or equivalently
\begin{equation}
h_{xy}^{2}(k)=[E_{\pm}(k)-h_{z}(k)]^{2}.
\label{Eq:CriticalK2}
\end{equation}
This expression can be simplified to $h_{z}(k)=\omega/2$, which entails the critical condition
\begin{equation}
\cos k_{c}=\frac{\omega-\delta_{2}}{\delta_{1}}.
\label{Eq:kc}
\end{equation}
This condition can only be satisfied if
\begin{equation}
\left|\frac{\omega-\delta_{2}}{\delta_{1}}\right|\leq1.
\label{Eq:CriticalK3}
\end{equation}
Therefore, there are Floquet DQPTs at critical times $t_{\pm}=(2n-1)T/2$ for all $n\in\mathbb{Z}^{+}$ if the condition (\ref{Eq:CriticalK3}) is satisfied.

To understand the physical meaning of this condition, we first consider
two limiting cases. In high-frequency limit ($\omega\rightarrow\infty$),
the condition (\ref{Eq:CriticalK3}) can not be satisfied for finite values of $\delta_{1}$
and $\delta_{2}$. So there is no Floquet DQPTs in high-frequency
limit. In low frequency limit ($\omega\rightarrow0$), there are Floquet
DQPTs if $|\delta_{2}/\delta_{1}|\leq1$. However, in this limit,
we also have $T\rightarrow\infty$ and we need to wait an infinite
long time for the first Floquet DQPT to happen. So practically there is also
no Floquet DQPTs in low frequency limit in our model. Putting together,
nontrivial regimes for Floquet DQPTs to happen appear
only at finite driving frequencies, independent of the driving amplitude
$\Omega/2$.

To further understand Eq.~(\ref{Eq:CriticalK3}), we recall the Floquet effective Hamiltonian $H_{F}(k)$ in Eq.~\ref{Eq:staticH}.
This Hamiltonian defines a two-component vector $[\Omega\sin k,\delta_{2}-\omega+\delta_{1}\cos k]$, the trajectory of which versus $k$ forms an ellipse confined in the $x$-$z$ plane.
The center of the ellipse is located on the $z$-axis, and its two vertices are at $(\delta_{2}-\omega\pm\delta_{1},0)$.
The ellipse will encircle the origin $(0,0)$ of the $z$-$x$ plane if and only if
\begin{equation}
(\omega-\delta_{2}-\delta_{1})(\omega-\delta_{2}+\delta_{1})<0.
\label{Eq:EncirclingCond}
\end{equation}
The number of times that the vector $(\delta_{2}-\omega+\delta_{1}\cos k,\Omega\sin k)$
encircles the origin of $z$-$x$ plane as $k$ changes from $0$
to $2\pi$ defines a winding number, which is a topological
invariant. For $H_{F}(k)$, it equals to $1$ (counterclockwise
encircling) or $-1$ (clockwise encircling) when the condition (\ref{Eq:EncirclingCond})
is satisfied. Furthermore, Eq.~(\ref{Eq:EncirclingCond}) is equivalent to $|\omega-\delta_{2}|<|\delta_{1}|$,
which is nothing but Eq.~(\ref{Eq:CriticalK3}), so long as the spectrum of $H_F(k)$ is gapped.

So in our driven qubit model, we find that there are Floquet DQPTs if the effective Floquet Hamiltonian $H_{F}(k)$ is topologically nontrivial.
In the following subsections, we will further build the connection between Floquet DQPTs and the topological phases of our Floquet system with the help of the symmetry classification of Floquet topological states.

\subsubsection{4. Meaning of critical momentum}\label{Sec:kcMeaning}
To further understand the physical meaning of critical momentum $k_c$, let's focus on the evolution of the Bloch state right at $k_c$. According to Eq.~(\ref{Eq:kc}), the Floquet effective Hamiltonian at $k_c$ is
\begin{equation}
H_F(k_c)=\frac{\omega}{2}\sigma_0+\frac{\Omega\sin(k_c)}{2}\sigma_x.
\end{equation}
The resulting Floquet modes at $t=0$ are thus eigenstates of $\sigma_x$, i.e., $|\varphi_\pm(k_c,0)\rangle=\frac{1}{\sqrt{2}}(1,\pm1)^\top$.
The propagator evolving this state from $t=0$ to $t_c=T/2$ is given by
\begin{equation}
U(k_c,T/2)=U_R(T/2)e^{-iH_F(k_c)T/2}=e^{-i\frac{\pi}{2}\sigma_z}e^{-i\frac{\Omega T}{4}\sin k_c\sigma_x}=e^{-ih_f(k_c,T/2)T/2},
\end{equation}
where $h_f(k_c,T/2)$ is interpreted as an effective Hamiltonian for the evolution from $t=0$ to $t_c=T/2$.
It is clear that the initial state $|\varphi_\pm(k_c,0)\rangle$ is an equal-amplitude superposition of the two eigenstates of $h_f(k_c,T/2)$, and therefore populating equally its two levels. This generalizes the situation in conventional DQPTs~\cite{HeylPRL2013}, where a time-independent final Hamiltonian $h_f$ [analogous to our $h_f(k_c,T/2)$ here], and usually with more complicated structures, is obtained by performing a quench across a quantum critical point. The quench-free protocol and the simple structure of $h_f(k_c,T/2)$ simplify the simulation of Floquet DQPTs in our experimental platform and make the transitions more robust to observe in long-time domains.

\subsubsection{5. Floquet DQPTs and Floquet topological states}\label{Sec:FloquetDQPTvsTOP}
From Eq.~(\ref{GenSol}) one can see that the evolution operator at a given $k$ is
\begin{equation}
U(k,t)=U_{R}(t)e^{-iH_{F}(k)t}=e^{-i\frac{\omega}{2}t\sigma_{z}}e^{-i\left\{ \left[h_{z}(k)-\frac{\omega}{2}\right]\sigma_{z}+h_{xy}(k)\sigma_{x}\right\} t}.
\end{equation}
One can then introduce two symmetric time frames, in which the Floquet operators are
\begin{alignat}{1}
U_{1}(k)&=  e^{-i\left\{ \left[h_{z}(k)-\omega/2\right]\sigma_{z}+h_{xy}(k)\sigma_{x}\right\} {T}/{2}}e^{-i\pi\sigma_{z}}e^{-i\left\{ \left[h_{z}(k)-\omega/2\right]\sigma_{z}+h_{xy}(k)\sigma_{x}\right\} {T}/{2}}=  -e^{-i\left\{ \left[h_{z}(k)-\omega/2\right]\sigma_{z}+h_{xy}(k)\sigma_{x}\right\} T},\\
U_{2}(k)&=  e^{-i\frac{\pi}{2}\sigma_{z}}e^{-i\left\{ \left[h_{z}(k)-\omega/2\right]\sigma_{z}+h_{xy}(k)\sigma_{x}\right\} T}e^{-i\frac{\pi}{2}\sigma_{z}}=  -e^{-i\left\{ \left[h_{z}(k)-\omega/2\right]\sigma_{z}-h_{xy}(k)\sigma_{x}\right\} T}.
\end{alignat}
Both $U_{1}(k)$ and $U_{2}(k)$ have the chiral symmetry $\Gamma=\sigma_{y}$, in the sense that
\begin{equation}
\Gamma U_{1}(k)\Gamma=U_{1}^{\dagger}(k),\qquad\Gamma U_{2}(k)\Gamma=U_{2}^{\dagger}(k).
\end{equation}
They are both unitarily equivalent to $U(k,T)$.

Up to a global constant, the effective Hamiltonians in these two time
frames are given by
\begin{equation}
H_{F}^{(1)}(k)=\left[h_{z}(k)-\frac{\omega}{2}\right]\sigma_{z}+h_{xy}(k)\sigma_{x},
\end{equation}
\begin{equation}
H_{F}^{(2)}(k)=\left[h_{z}(k)-\frac{\omega}{2}\right]\sigma_{z}-h_{xy}(k)\sigma_{x}.
\end{equation}
Now let $W_{1}$ and $W_{2}$ be the winding numbers of $H_{F}^{(1)}(k)$
and $H_{F}^{(2)}(k)$, respectively. Using the components of $H_{F}^{(1)}(k)$ and $H_{F}^{(2)}(k)$, they can be written as
\begin{alignat}{1}
W_{1}= & +\int_{-\pi}^{\pi}\frac{dk}{2\pi}\frac{\left[h_{z}(k)-\omega/2\right]\partial_{k}h_{xy}(k)-h_{xy}(k)\partial_{k}\left[h_{z}(k)-\omega/2\right]}{\left[h_{z}(k)-\omega/2\right]^{2}+h_{xy}^{2}(k)},\\
W_{2}= & -\int_{-\pi}^{\pi}\frac{dk}{2\pi}\frac{\left[h_{z}(k)-\omega/2\right]\partial_{k}h_{xy}(k)-h_{xy}(k)\partial_{k}\left[h_{z}(k)-\omega/2\right]}{\left[h_{z}(k)-\omega/2\right]^{2}+h_{xy}^{2}(k)}.
\end{alignat}
It is clear that we always have $W_{2}=-W_{1}$. According to the topological classification
of chiral symmetric Floquet systems in one-dimension~\cite{AsbothPRB2013}, the Floquet
operator $U(k,T)$ can be characterized by a pair of winding numbers $W_0$ and $W_\pi$, given by
\begin{equation}
W_{0}=\frac{W_{1}+W_{2}}{2}=0,\qquad W_{\pi}=\frac{W_{1}-W_{2}}{2}=W_{1}.
\end{equation}
They count the number of zero and $\pi$ edge modes
under open boundary conditions of the lattice model~\cite{AsbothPRB2013}.

The topological nontrivial regime of $U(k,T)$ is given by Eq.~(\ref{Eq:CriticalK3}), where we have $W_{1}=1$. Therefore, in this regime the system described by $U(k,T)$ is in a topologically nontrivial phase with winding numbers $W_{0}=0$ and $W_{\pi}=1$, possessing a pair of $\pi$ quasienergy edge modes under open boundary conditions. Surprisingly, Eq.~(\ref{Eq:CriticalK3}) is also the condition under which Floquet DQPTs happen. We therefore unveil an explicit connection between the existence of Floquet DQPTs in micromotion times and the topological nature of the underlying stroboscopic Floquet states. In the following subsections, we will further characterize the topological features of Floquet DQPTs by a dynamical topological invariant.

\subsubsection{6. Geometric phase and dynamical topological invariant}
The geometric phase of return amplitude (\ref{RTAmp}) can be used to construct a
dynamical topological invariant for characterizing Floquet DQPTs. This non-adiabatic, non-cyclic
geometric phase is given by
\begin{equation}
\phi_{\pm}^{{\rm geo}}(k,t)=\phi_{\pm}(k,t)-\phi_{\pm}^{{\rm dyn}}(k,t),
\end{equation}
where the total phase
\begin{equation}
\phi_{\pm}(k,t)=-i\ln\left[\frac{G_{\pm}(k,t)}{|G_{\pm}(k,t)|}\right],
\end{equation}
and the dynamical phase
\begin{equation}
\phi_{\pm}^{{\rm dyn}}(k,t)=-\int_{0}^{t}ds\langle\psi_{\pm}(k,s)|H(k,s)|\psi_{\pm}(k,s)\rangle=-\langle\varphi_{\pm}(k)|H_{R}(k)|\varphi_{\pm}(k)\rangle\cdot t,
\end{equation}
with $H_{R}(k)=U_{R}^{\dagger}(t)H(k,t)U_{R}(t)$.
For our driven qubit model, the geometric meaning of $\phi_{\pm}^{{\rm geo}}(k,t)$
is the area subtended by a solid angle on the Bloch sphere. The boundary
of the area contains two parts. One of them is the physical trajectory
of the system, and the other one is the geodesic connecting two open
ends of the physical trajectory~\cite{GeoPhas}. As shown in the main text, the pattern of $\phi_{\pm}^{{\rm geo}}(k,t)$
versus $k$ and $t$ could help us to find signatures of Floquet DQPTs.

As an example, consider the behavior of geometric phase $\phi_{\pm}^{{\rm geo}}(k_{c},t)$
at the critical momentum $k_{c}$~\cite{HeylPRB2016}. In this case,
the total phase is

\begin{alignat*}{1}
\phi_{\pm}(k_{c},t)= & -E_{\pm}(k_{c})t-i\ln\left[\frac{h_{xy}^{2}(k_{c})+e^{i\omega t}[E_{\pm}(k_{c})-h_{z}(k_{c})]^{2}}{|h_{xy}^{2}(k_{c})+e^{i\omega t}[E_{\pm}(k_{c})-h_{z}(k_{c})]^{2}|}\right]\\
= & -E_{\pm}(k_{c})t-i\ln\left[\frac{1+e^{i\omega t}}{|1+e^{i\omega t}|}\right]=-E_{\pm}(k_{c})t+\frac{\omega t}{2}-i\ln\left[\frac{\cos(\pi t/T)}{|\cos(\pi t/T)|}\right],
\end{alignat*}
and the dynamical phase is
$$
\phi_{\pm}^{{\rm dyn}}(k,t)= -\langle\varphi_{\pm}(k_{c})|H_{R}(k_{c})|\varphi_{\pm}(k_{c})\rangle t=  -E_{\pm}(k_{c})t+\frac{[E_{\pm}(k_{c})-h_{z}(k_{c})]^{2}}{h_{xy}^{2}(k_{c})+[E_{\pm}(k_{c})-h_{z}(k_{c})]^{2}}\omega t=-E_{\pm}(k_{c})t+\frac{\omega t}{2}.
$$
So the geometric phase is
\[
\phi_{\pm}^{{\rm geo}}(k_{c},t)=\phi_{\pm}(k_{c},t)-\phi_{\pm}^{{\rm dyn}}(k_{c},t)=-i\ln\{{\rm sgn}[\cos(\pi t/T)]\},
\]
and it is straightforward to see that
\[
\cos(\pi t/T)\begin{cases}
>0 & nT<t<(n+1/2)T\\
<0 & (n+1/2)T<t<(n+1)T
\end{cases}\qquad\forall n\in\mathbb{Z}^{+}.
\]
Therefore the geometric phase $\phi_{\pm}^{{\rm geo}}(k_{c},t)$ has
a $\pi$ jump at the midpoint of each driving period. Note that this
will not happen if there is no solution of critical momentum $k_{c}\in(0,\pi)$.
Such a discontinuous behavior of the geometric phase at the critical
momentum suggests a topological origin of Floquet DQPTs.

Using the geometric phase $\phi_{\pm}^{{\rm geo}}(k,t)$, we can construct a time-dependent dynamical topological invariant
\begin{equation}
\nu_{\pm}(t)=\int_{0}^{\pi}\frac{dk}{2\pi}\left[\partial_{k}\phi_{\pm}^{{\rm geo}}(k,t)\right].
\end{equation}
At a given time $t$, it counts the number of times the geometric
phase $\phi_{\pm}^{{\rm geo}}(k,t)$ winds around the Brillouin
zone. When the evolution of the system passes the critical time of
a Floquet DQPT, the value of $\nu_{\pm}(t)$ will gets a quantized
jump if the condition (\ref{Eq:CriticalK3}) is satisfied. Therefore, the winding number
$\nu_{\pm}(t)$ establishes a direct connection between the
topological nature of Floquet DQPTs in our system and the topological
property of the evolving Floquet states.

\subsection{F. Numerical examples}

In previous subsections, we have introduced relevant quantities to
characterize the Floquet DQPTs in a driven spin chain. In this subsection,
we present numerical results to support our theoretical findings.
Both situations with and without Floquet DQPTs will be considered.

\begin{figure}\centering
	\includegraphics[scale=0.5]{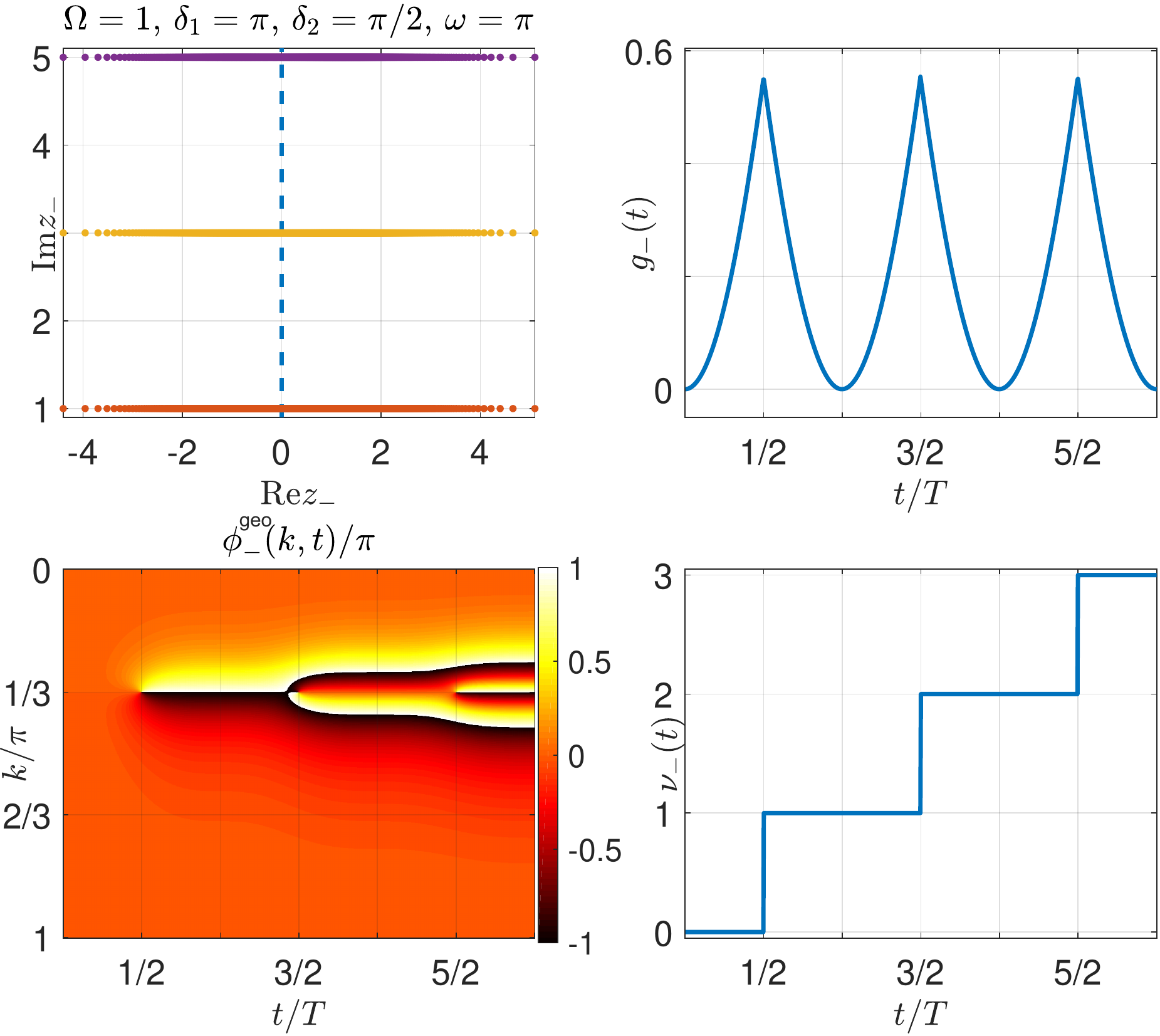}
	\caption{Lines of Fisher zeros, rate function of return probability, geometric phase and dynamical topological invariant of the driven spin chain model. System parameters are chosen as $\Omega=1$, $\delta_{1}=\pi$, $\delta_{2}=\pi/2$, and $\omega=\pi$. The initial state fills the lower Floquet band.}
	\label{Fig:FloquetDQPTExp1}
\end{figure}

In the first example as shown in Fig.~\ref{Fig:FloquetDQPTExp1}, we choose the system parameters to be $\Omega=1$,
$\delta_{1}=\pi$, $\delta_{2}=\pi/2$, $\omega=\pi$, and the initial
state to be the Floquet modes $\prod_{k>0}|\varphi_{-}(k,t)\rangle$
filling the lower Floquet band. It is directly seen that $|\omega-\delta_{2}|/|\delta_{1}|=1/2<1$.
So there will be Floquet DQPTs at critical times $t_{-}=\frac{2n-1}{2}T=2n-1$,
and the critical momentum is $k_{c}=\pi/3$. The figure above shows
numerical calculation of relevant quantities over three driving periods.
We see that each line of Fisher zeros crosses the imaginary time axis
at a given critical time. At each of the critical time, the rate function
of return probability $g_{-}(t)$ behaves non-analytically as a function
of $t$, characterized by a kink structure around each $t_{c}$ and
repeating at $t_{c}+nT$ for any $n\in\mathbb{Z}$. Similarly, the
number of $2\pi$-jumps in the pattern of geometric phase $\phi_{-}^{{\rm geo}}(k,t)$
versus $k$ increases by $1$ each time when the dynamics passes through
a critical time. This is clearly reflected in the behavior of winding
number $\nu_{-}(t)$ in time.

\begin{figure}\centering
	\includegraphics[scale=0.5]{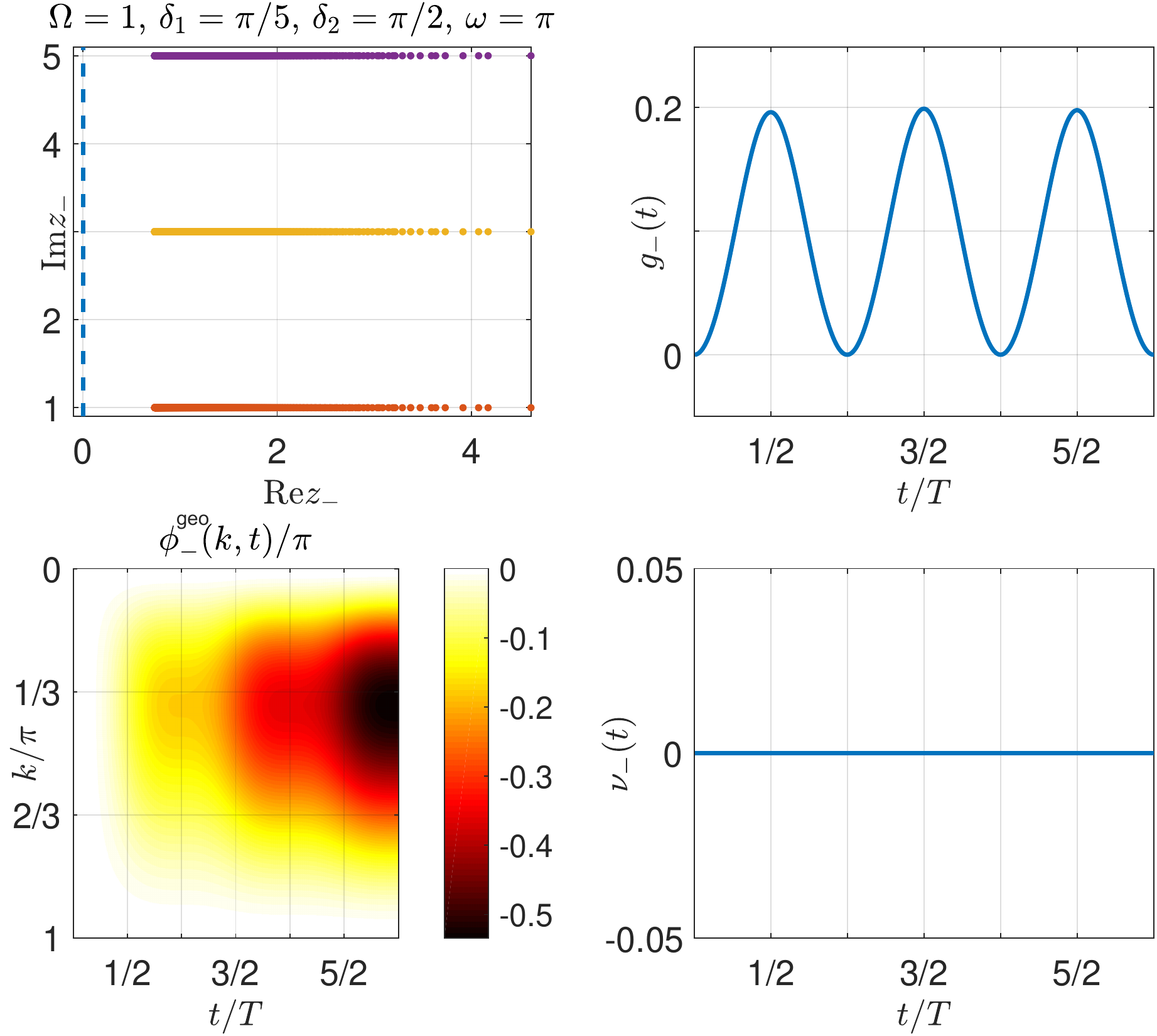}
	\caption{Lines of Fisher zeros, rate function of return probability, geometric phase and dynamical topological invariant of the driven spin chain model. System parameters are chosen as $\Omega=1$, $\delta_{1}=\pi/5$, $\delta_{2}=\pi/2$, and $\omega=\pi$. The initial state fills the lower Floquet band.}
	\label{Fig:FloquetDQPTExp2}
\end{figure}

In the second example as shown in Fig.~\ref{Fig:FloquetDQPTExp2}, we choose the system parameters to be $\Omega=1$,
$\delta_{1}=\pi/5$, $\delta_{2}=\pi/2$, $\omega=\pi$, and the initial
state still to be the Floquet modes $\prod_{k>0}|\varphi_{-}(k,t)\rangle$
filling the lower Floquet band. This time we have $|\omega-\delta_{2}|/|\delta_{1}|=5/2>1$.
So the condition (\ref{Eq:CriticalK3}) is not satisfied and there will be no Floquet
DQPTs. The numerical calculations of Fisher zeros, rate function of
return probability, geometric phase and winding number in three driving
periods as presented in Fig.~\ref{Fig:FloquetDQPTExp2} clearly justify this finding.

\begin{figure}\centering
	\includegraphics[scale=0.5]{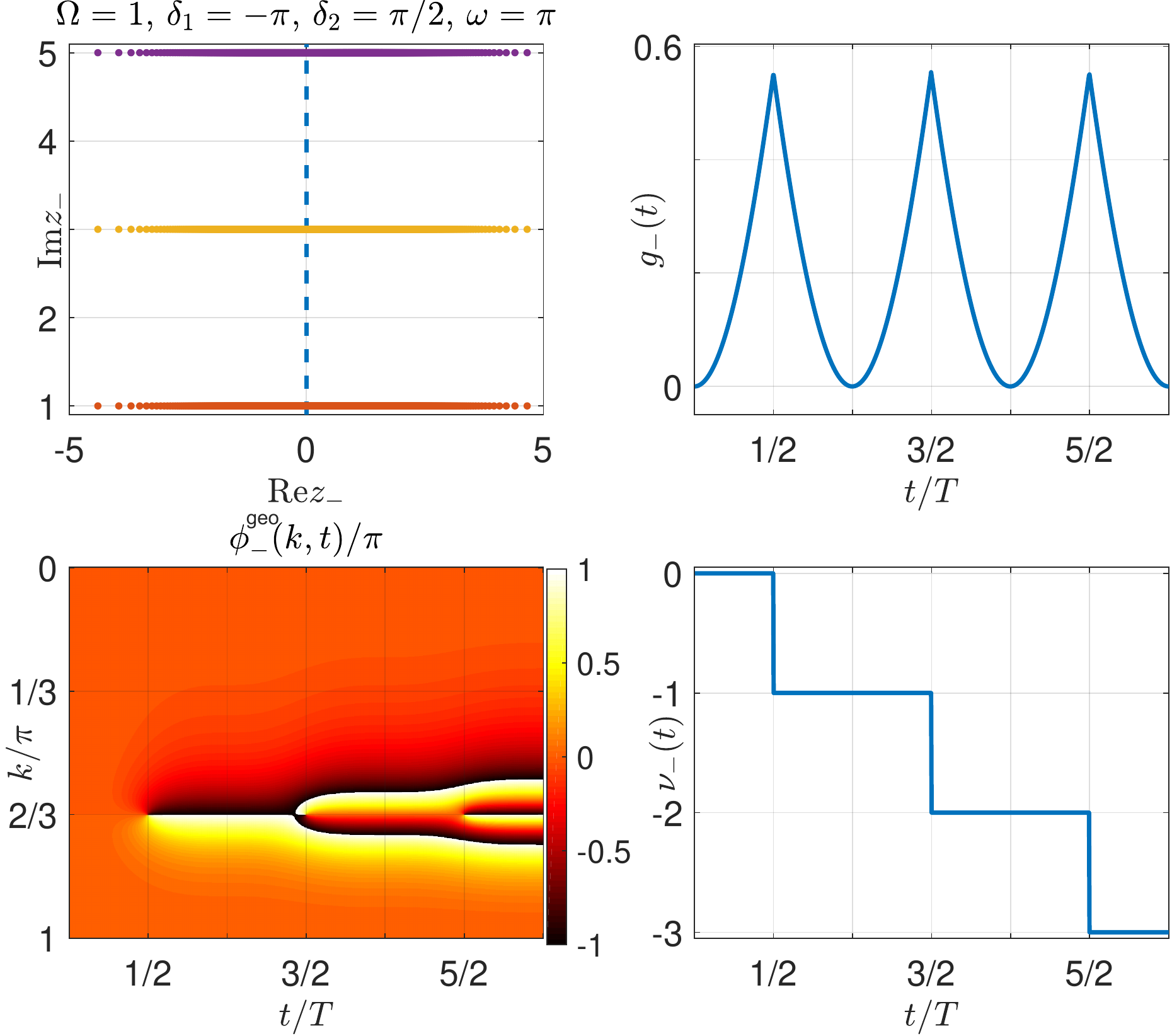}
	\caption{Lines of Fisher zeros, rate function of return probability, geometric phase and dynamical topological invariant of the driven spin chain model. System parameters are chosen as $\Omega=1$, $\delta_{1}=-\pi$, $\delta_{2}=\pi/2$, and $\omega=\pi$. The initial state fills the lower Floquet band.}
	\label{Fig:FloquetDQPTExp3}
\end{figure}

As a final example shown in Fig.~\ref{Fig:FloquetDQPTExp3}, we choose the system parameters to be $\Omega=1$,
$\delta_{1}=-\pi$, $\delta_{2}=\pi/2$, $\omega=\pi$, and the initial
state to be the Floquet modes $\prod_{k>0}|\varphi_{-}(k,t)\rangle$
filling the lower Floquet band. It is seen that $|\omega-\delta_{2}|/|\delta_{1}|=1/2<1$.
So there will be Floquet DQPTs at critical times $t_{-}=\frac{2n-1}{2}T=2n-1$,
and the critical momentum is $k_{c}=\frac{2\pi}{3}$. The figure above
shows numerical calculation of relevant quantities over three driving
periods. We see that each line of Fisher zeros crosses the imaginary
time axis at a given critical time. At each of the critical time,
the rate function of return probability $g_{-}(t)$ behaves non-analytically
as a function of $t$, characterized by a kink structure around each
$t_{c}$ and repeating at $t_{c}+nT$ for any $n\in\mathbb{Z}$. Similarly,
the number of $2\pi$-jumps in the pattern of geometric phase $\phi_{-}^{{\rm geo}}(k,t)$
versus $k$ increases by $1$ each time when the dynamics passes through
a critical time. This is clearly reflected in the behavior of winding
number $\nu_{-}(t)$ in time.

\section{II. Experiment}

\subsection{A. Experimental setup}
The experiment is performed on an NV center in a \{100\}-face bulk diamond synthesized by chemical vapor deposition (CVD).
The nitrogen impurity is less than 5 ppb and the abundance of $^{13}$C is at the natural level of about 1.1\%. The dephasing time of the NV electron spin is 1.7 $\mu$s.
The NV center is optically addressed by a home-built confocal microscope. Green laser is used for optical excitation.
The laser beam is released and cut off by an acousto-optic modulator (power leakage ratio $\sim1/1000$). To reduce the laser leakage further, the beam passes twice through the acousto-optic modulator. The laser is focused into the diamond by an oil objective (60*O, NA 1.42). The phonon sideband fluorescence with the wavelength between 650 and 800 nm is collected by the same oil objective and finally detected by an avalanche photodiode with a counter card. A solid immersion lens etched on the diamond by focused ion beam enhances the fluorescence counting rate up to 400 thousand counts per second.
The microwaves are generated by an arbitrary waveform generator (AWG) and then strengthened by a power amplifier. Finally, the microwaves are radiated to the NV center from a coplanar waveguide. The magnetic field is supplied by a permanent magnet mounted on a translation stage.

\subsection{B. Calibration}
The transverse component of the rotating field is $\omega\sin k$. In order to feed the microwaves with proper amplitude to the NV center, we calibrate the Rabi frequency as a function of the AWG's output amplitude. The calibration is done by performing conventional Rabi oscillation experiments with various output amplitudes. We fit the experimental data of Rabi oscillation associated with each output amplitude $V$ to extract the corresponding Rabi frequency $\omega_{\rm R}$, and then fit the Rabi frequency using $\omega_{\rm R}=a \exp(-V^2/b^2)V$ with $a$ and $b$ being the coefficients to be determined. The relation between the Rabi frequency and the AWG's output amplitude is thus obtained. Such calibration is carried out hourly to guard against the drift of experimental conditions.

\subsection{C. Pulse sequence}
After the qubit is polarized to the state $\left| {{\psi _0}} \right\rangle={(1,0)^{\rm{T}}}$ by a green laser pulse, a resonant microwave pulse is applied to prepare the initial state.
In the laboratory frame, the Hamiltonian of the qubit irradiated by the pulse is
\begin{equation}\label{PrepLab}
H_{{\rm{ini}}}^{{\rm{lab}}} = \frac{\omega _0}{2}\sigma _z + \omega_1 \cos ( {\omega _0} t +\varphi_{\rm{ini}})\sigma _x.
\end{equation}
where the first term on the right-hand side is the static component of the Hamiltonian with $\omega _0$ being the resonant frequency,
and the second term accounts for the effect of the microwaves with $\omega_1$, $\varphi_{\rm{ini}}$, and $t$ being the Rabi frequency, the initial phase, and the time starting from zero, respectively.
% $\omega_1$ is the Rabi frequency, $\varphi_{\rm{ini}}= {\rm sgn}({\delta _1}\cos \theta  + {\delta _2}){\pi}/{2}$ is the initial phase of the microwave, and $t$ is the time starting from zero.
The initial phase $\varphi_{\rm{ini}}$ is set as $-\pi/2$ in this experiment.
%The first term of the right-hand side accounts for the microwave, and the second term accounts for the static energy splitting.
In the rotating frame that rotates around the $z$ axis with the angular frequency $\omega_0$, or to put it another way, under the rotating transformation characterized by the rotation operator $R_{\rm{ini}} = {e^{ - i{\omega _0} t{\sigma _z}/2}}$,
the Hamiltonian in Eq.~(\ref{PrepLab}) is transformed to
\begin{equation}\label{PrepRot}
H_{{\rm{ini}}}^{{\rm{rot}}} = {R_{\rm{ini}}^\dag }H_{{\rm{ini}}}^{{\rm{lab}}}R_{\rm{ini}} - i{R_{\rm{ini}}^\dag }\frac{d}{{dt}}R_{\rm{ini}} = \frac{\omega _1}{2} ( \sigma _x\cos \varphi_{\rm{ini}} +\sigma _y\sin \varphi_{\rm{ini}}),
\end{equation}
where the second equality relies on the rotating wave approximation.
The pulse lasts for $t_{\rm{ini}}=\beta/\omega_1$, where $\beta$ is the inclination angle of the initial state, namely,
\begin{equation}\label{IniPolarAngle}
\beta=\pi-\arccos\frac{ \delta_1 \cos k +\delta_2 -\omega }{{\sqrt {{{(\Omega \sin k )}^2} + {{({\delta _1}\cos k  + {\delta _2}-\omega)}^2}} }}.
\end{equation}
After the pulse, the state of the qubit in the rotating frame is
\begin{equation}\label{IniStateRot}
\left| {\psi _{{\rm{ini}}}^{{\rm{rot}}}} \right\rangle = e^{-iH_{{\rm{ini}}}^{{\rm{rot}}}t_{\rm{ini}}} \left| {{\psi _0}} \right\rangle = \left| {\psi {\rm{(}}k,0{\rm{)}}} \right\rangle,
\end{equation}
which is our desired initial state. In the laboratory frame, this state immediately after the pulse is expressed as
\begin{equation}\label{IniStateLab}
\left| {\psi _{{\rm{ini}}}^{{\rm{lab}}}} \right\rangle  = {e^{ - i{\omega _0} t_{\rm{ini}} {\sigma _z}/2}} \left| {\psi _{{\rm{ini}}}^{{\rm{rot}}}} \right\rangle.
\end{equation}

Next, a microwave pulse is applied to build the model Hamiltonian in Eq.~(\ref{eq:Hqubit}). In most cases, this pulse is off-resonant.
In the laboratory frame, the Hamiltonian of the qubit irradiated by the pulse is
\begin{equation}\label{PumpLab}
H_{{\rm{evo}}}^{{\rm{lab}}} = \frac{\omega _0}{2}\sigma _z + \Omega \sin k \cos \left[ ({\omega _0} - {\delta _1}\cos k  - {\delta _2}+\omega)t + \omega _0 t_{\rm{ini}}\right]\sigma _x,
\end{equation}
where $t$ is the time starting from zero.
%Note that, in the preceding sections of the text and the main text, we adopt the scaled time $\tau=t/T$. But here we use the real time $t$ when discussing the microwave pulse.
In the rotating frame with the rotation operator
\begin{equation}
R_{\rm{evo}} = {e^{ - i[({\omega _0} - {\delta _1}\cos k  - {\delta _2})t + \omega _0 t_{\rm{ini}}]{\sigma _z}/2}},
\end{equation}
the Hamiltonian in Eq.~(\ref{PumpLab}) is transformed to our target Hamiltonian, namely,
\begin{equation}\label{PumpRot}
H_{{\rm{evo}}}^{{\rm{rot}}} = {R_{\rm{evo}}^\dag }H_{{\rm{evo}}}^{{\rm{lab}}}R_{\rm{evo}} - i{R_{\rm{evo}}^\dag }\frac{d}{{dt}}R_{\rm{evo}}
= \frac{\Omega \sin k}{2} \left[ {\cos (\omega t) {{\sigma _x}} + \sin (\omega t) {{\sigma _y}}} \right]+\frac{{\delta _1}\cos k+{\delta _2}}{2}\sigma _z,
\end{equation}
where the second equality relies on the rotating wave approximation.
In this rotating frame, the state in Eq.~(\ref{IniStateLab}) is rewritten as
\begin{equation}\label{IniStateRot}
R_{\rm{evo}}^\dag(t=0)\left| {\psi _{{\rm{ini}}}^{{\rm{lab}}}} \right\rangle=\left| {\psi _{{\rm{ini}}}^{{\rm{rot}}}} \right\rangle  = \left| {\psi {\rm{(}}k,0{\rm{)}}} \right\rangle,
\end{equation}
%which is just our desired one.
which has the same form as in the rotating frame defined by $R_{\rm{ini}}$.
The pulse lasts for some duration duration $t_{\rm e}$, which is a sampling point in time.
Assume that the state of the qubit immediately after the pulse is $\left| {\psi _{{\rm{pump}}}^{{\rm{rot}}}} \right\rangle$ in this rotating frame.
In the laboratory frame, the state is expressed as
\begin{equation}\label{PumpStateLab}
\left| {\psi _{{\rm{evo}}}^{{\rm{lab}}}} \right\rangle={e^{ - i[({\omega _0} - {\delta _1}\cos k  - {\delta _2})t_{\rm e} + \omega _0 t_{\rm{ini}}]{\sigma _z}/2}}\left| {\psi _{{\rm{evo}}}^{{\rm{rot}}}} \right\rangle.
\end{equation}

Finally, a resonant microwave pulse is applied to assist measurement.
In the laboratory frame, the Hamiltonian of the qubit irradiated by the pulse is
\begin{equation}\label{FinLab}
H_{{\rm{fin}}}^{{\rm{lab}}} = \frac{\omega _0}{2}\sigma _z + \omega_1 \cos \left[ {{\omega _0}t + ({\omega _0} - {\delta _1}\cos k  - {\delta _2})t_{\rm e} } + \omega _0 t_{\rm{ini}} + \varphi_{\rm{fin}}\right]\sigma _x,
\end{equation}
with $\varphi_{\rm{fin}}=\pi/2$. Here $t$ is also the time starting from zero.
In the rotating frame with the rotation operator
\begin{equation}
R_{\rm{fin}} = {e^{ - i[{\omega _0} t+({\omega _0} - {\delta _1}\cos k  - {\delta _2})t_{\rm e}+\omega _0 t_{\rm{ini}}]{\sigma _z}/2}},
\end{equation}
the Hamiltonian in Eq.~(\ref{FinLab}) is transformed to
\begin{equation}\label{ReadRot}
H_{{\rm{fin}}}^{{\rm{rot}}} = {R_{\rm{fin}}^\dag }H_{{\rm{fin}}}^{{\rm{lab}}}R_{\rm{fin}} - i{R_{\rm{fin}}^\dag }\frac{d}{{dt}}R_{\rm{fin}}
= \frac{\omega _1}{2} (\sigma _x\cos \varphi_{\rm{fin}} +\sigma _y\sin \varphi_{\rm{fin}}),
\end{equation}
where the second equality is based on the rotating wave approximation.
In this rotating frame, the state in Eq.~(\ref{PumpStateLab}) is rewritten as
\begin{equation}\label{PumpStateRot}
R_{\rm{fin}}^\dagger(t=0)\left| {\psi _{{\rm{evo}}}^{{\rm{lab}}}} \right\rangle = \left| {\psi _{{\rm{evo}}}^{{\rm{rot}}}} \right\rangle,
\end{equation}
which has the same form as in the rotating frame defined by $R_{\rm{evo}}$.
The pulse lasts for $t_{\rm{ini}}=\beta/\omega_1$.
After the pulse, laser illumination is carried out to measure the probability in $| 0 \rangle$.
The combined effect of the final microwave pulse and its subsequent laser illumination amounts to the measurement of
\begin{equation}\label{VelMeas}
e^{iH_{{\rm{fin}}}^{{\rm{rot}}}t_{\rm{fin}}} |0\rangle \langle0| e^{-iH_{{\rm{fin}}}^{{\rm{rot}}}t_{\rm{fin}}}= |\psi(k,0)\rangle \langle\psi(k,0)|,
\end{equation}
which yields the return probability $|\langle \psi(k,0) | \psi(k,t) \rangle|^2$.
%All the microwave used in this experiment is generated by an AWG.

\subsection{D. Experimental data analysis}
As shown in Fig.~2(c) of the main text, the spin state is read out during the latter laser pulse and there are two counting windows. Such sequence is iterated five hundred thousand times. The total photon count recorded by the first (second) window during these iterations is regarded as signal (reference) and denoted by $s$ ($r$). The raw experimental data is $x=s/r$. To normalize the data, a conventional Rabi oscillation is performed alongside. We fit the raw data of the Rabi oscillation using the function $x=x_0+a\cos(\omega_{\rm R} t+\varphi)$, and then normalize the experimental data as $x_{\rm{n}}=(x-x_0)/(2a)+1/2$. The data thus normalized represent the probability in $| 0 \rangle$. In the experiment, the sampling points in $t$ are 0, 0.020, 0.040, 0.060, 0.080, 0.094, 0.096, 0.098, 0.100, 0.102, 0.104, 0.106, 0.120, 0.140, 0.160, 0.180, 0.200, 0.220, 0.240, 0.260, 0.280, 0.294, 0.296, 0.298, 0.300, 0.302, 0.304, 0.306, 0.320, 0.340, 0.360, 0.380, 0.400, 0.420, 0.440, 0.460, 0.480, 0.494, 0.496, 0.498, 0.500, 0.502, 0.504, 0.506, 0.520, 0.540, 0.560, 0.580, and 0.600 $\mu$s.
In the experiment data illustrated in Figs.~3 and 4(b) of the main text and Figs.~\ref{retprob} and \ref{expectation}(d)-(f) of this supplemental material, the sampling interval in $k$ is $\pi/12$.
In the experiment data illustrated in Figs.~4(a) of the main text and Figs.~\ref{expectation}(a)-(c), the sampling intervals in $k$ are $\pi/180$.
The numerical integration over $k$ is based on the trapezoidal rule.

%\subsection{Method for measuring the geometric phase and dynamical topological invariant}
\subsection{E. Method for measuring the geometric phase}
The total phase, dynamic phase, and geometric phase, in turn, are
\begin{equation}\label{phases}
\begin{aligned}
 &{\phi_{\alpha}^{{\rm{tot}}}}(k,t) = \arg \left\langle {{\psi_{\alpha} (k,0)}}
 \mathrel{\left | {\vphantom {{\psi_{\alpha} (k,0)} {\psi_{\alpha} (k,t)}}}
 \right. \kern-\nulldelimiterspace}
 {{\psi_{\alpha} (k,t)}} \right\rangle  =  - {E_{\alpha} }t - \frac{\omega t}{2} + \arg \left\langle {{\varphi_{\alpha} (k,0)}}
 \mathrel{\left | {\vphantom {{\varphi_{\alpha} (k,0)} {\varphi_{\alpha} (k,t)}}}
 \right. \kern-\nulldelimiterspace}
 {{\varphi_{\alpha} (k,t)}} \right\rangle + 2 \pi n,  \\
 &{\phi_{\alpha}^{{\rm{dyn}}}}(k,t) =  - \int_0^t {ds\left\langle {\psi_{\alpha} (k,s)} \right|H(k,s)\left| {\psi_{\alpha} (k,s)} \right\rangle }  =  - {E_ \pm }t - \int_0^t {ds\frac{\omega }{2}\left\langle {\varphi_{\alpha} (k,s)} \right|{\sigma _z}\left| {\varphi_{\alpha} (k,s)} \right\rangle }.  \\
 &{\phi_{\alpha}^{{\rm{geo}}}}(k,t) ={\phi _{{\rm{tot}}}}(k,t) -{\phi _{{\rm{dyn}}}}(k,t)= \arg \langle \varphi_{\alpha}(k,0) | \varphi_{\alpha}(k,t) \rangle +\int_0^t {ds\frac{\omega }{2}\left\langle {\varphi_{\alpha} (k,s)} \right|{\sigma _z}\left| {\varphi_{\alpha} (k,s)} \right\rangle } - \frac{\omega t}{2} + 2\pi n,
\end{aligned}
\end{equation}
where $n$ is an integer.
After measuring
\begin{equation}\label{tomo}
\begin{aligned}
&\langle\sigma_x(k,t)\rangle :=\langle \psi_{\alpha} (k,t)| \sigma_x |\psi_{\alpha} (k,t)\rangle =\langle \varphi_{\alpha} (k,t)| \sigma_x |\varphi_{\alpha} (k,t)\rangle,\\
&\langle\sigma_y(k,t)\rangle :=\langle \psi_{\alpha} (k,t)| \sigma_y |\psi_{\alpha} (k,t)\rangle =\langle \varphi_{\alpha} (k,t)| \sigma_y |\varphi_{\alpha} (k,t)\rangle,\\
&\langle\sigma_z(k,t)\rangle :=\langle \psi_{\alpha} (k,t)| \sigma_z |\psi_{\alpha} (k,t)\rangle =\langle \varphi_{\alpha} (k,t)| \sigma_z |\varphi_{\alpha} (k,t)\rangle,
\end{aligned}
\end{equation}
the inner product $\langle \varphi_{\alpha}(k,0) | \varphi_{\alpha}(k,t) \rangle$ can be calculated. In our experiment, $\alpha=-$ is adopted and we have
\begin{equation}\label{product}
\langle \varphi_-(k,0) | \varphi_- (k,t) \rangle=\sin \frac{\theta }{2}\sqrt {\frac{{1 + \cos \vartheta }}{2}}  - {e^{i\phi}}\cos \frac{\theta }{2}\sqrt {\frac{{1 - \cos \vartheta }}{2}},
\end{equation}
where
\begin{equation}\label{angles}
\begin{aligned}
 &\theta  = \arccos \frac{{{\delta _1}\cos k + {\delta _2} - \omega }}{{\sqrt {{{({\delta _1}\cos k + {\delta _2} - \omega )}^2} + {{(\Omega \sin k)}^2}} }} \\
 &\vartheta  = \arccos \frac{{\left\langle {{\sigma _z}(k,t)} \right\rangle }}{{\sqrt {{{\left\langle {{\sigma _x}(k,t)} \right\rangle }^2} + {{\left\langle {{\sigma _y}(k,t)} \right\rangle }^2} + {{\left\langle {{\sigma _z}(k,t)} \right\rangle }^2}} }} \\
 &\phi  = \left\{ {\begin{array}{*{20}{c}}
   {\arccos \frac{{\left\langle {{\sigma _x}(k,t)} \right\rangle }}{{\sqrt {{{\left\langle {{\sigma _x}(k,t)} \right\rangle }^2} + {{\left\langle {{\sigma _y}(k,t)} \right\rangle }^2}} }},~~~{\rm{if}}~~~\left\langle {{\sigma _y}(k,t)} \right\rangle  \ge 0},  \\
   { - \arccos \frac{{\left\langle {{\sigma _x}(k,t)} \right\rangle }}{{\sqrt {{{\left\langle {{\sigma _x}(k,t)} \right\rangle }^2} + {{\left\langle {{\sigma _y}(k,t)} \right\rangle }^2}} }},~~~{\rm{if}}~~~\left\langle {{\sigma _y}(k,t)} \right\rangle  < 0}.  \\
\end{array}} \right.
\end{aligned}
\end{equation}
The geometric phase can then be acquired according to the last expression in Eq.~(\ref{phases}).
In this work, the geometric phase is set between $-\pi$ and $\pi$.

%The dynamical topological invariant can be calculated as
%\begin{equation}\label{angles}
%\begin{aligned}
%{\nu _\alpha }(t) = \int_0^\pi  {\frac{{dk}}{{2\pi }}\frac{{\partial \phi _\alpha ^{{\rm{geo}}}}}{{\partial k}}}  = \frac{1}{{2\pi }}\big\{ {\phi _\alpha ^{{\rm{geo}}}(\pi ,t) - \phi _\alpha ^{{\rm{geo}}}(0,t) + \sum\nolimits_j {\left[ {\phi _\alpha ^{{\rm{geo}}}(k_j^ - ,t) - \phi _\alpha ^{{\rm{geo}}}(k_j^ + ,t)} \right]} } \big\}.
%\end{aligned}
%\end{equation}
%Here $k_j$ denotes the discontinuity point of $\phi _\alpha ^{{\rm{geo}}}$ in the $k$ dimension, and its superscript $+$ or $-$ denotes the left-sided or right-sided limit, respectively.
%Because of the relation $\phi _\alpha ^{{\rm{geo}}}(\pi ,t) = \phi _\alpha ^{{\rm{geo}}}(0,t)$, one can obtain the dynamical topological invariant by evaluating the jumping ranges at discontinuity points.
%In the experiment, in order to discern the discontinuity points, we stipulate that, if the experimental geometric phase difference between two adjacent points in the $k$ dimension is above a threshold, a discontinuity point will be taken into account.
%% To be specific, let $k_{\rm I}$ and $k_{\rm II}$ denote a pair of adjacent points in the experimental data. We

\subsection{F. Experimental data}
\begin{figure}\centering
\includegraphics[width=1\columnwidth]{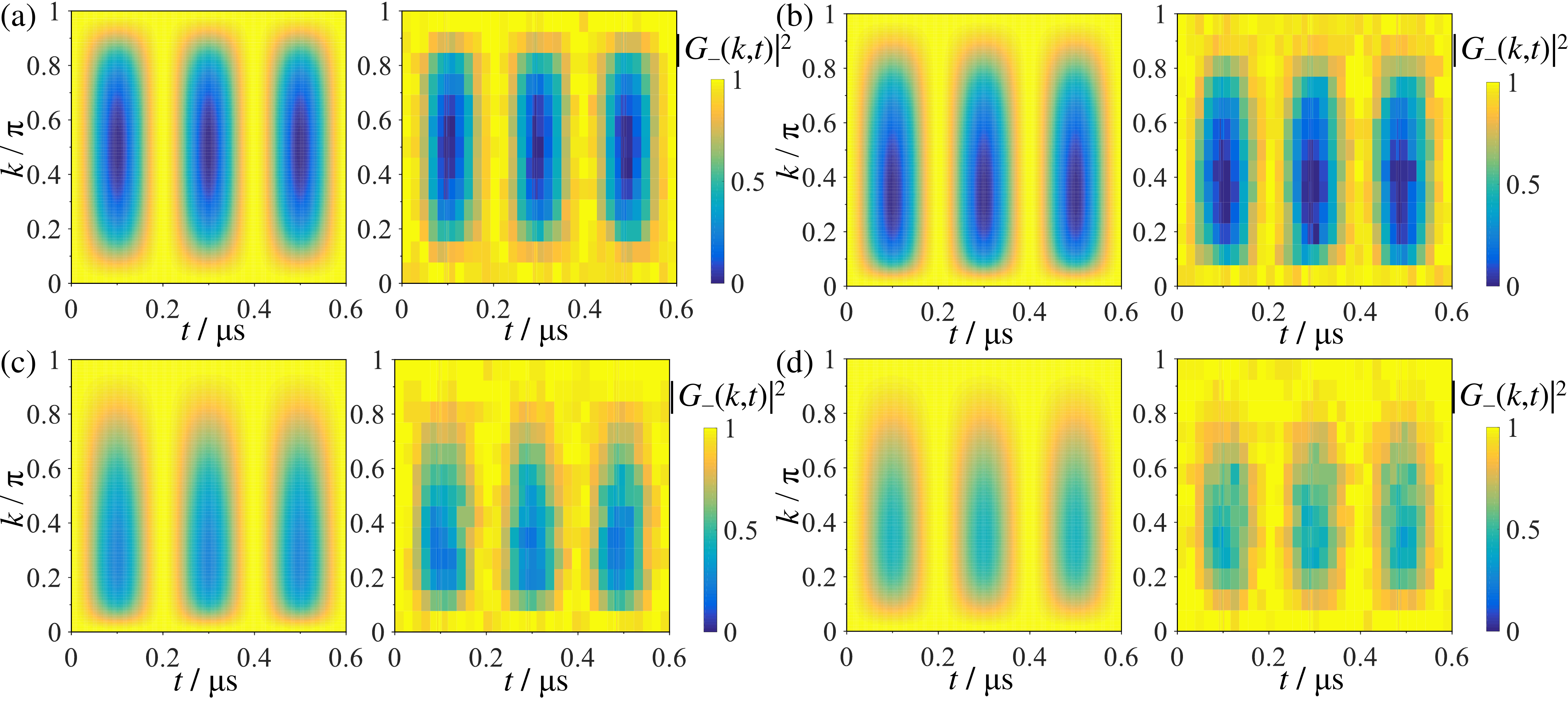}
\caption{Return probabilities.
   (a)-(d) Return probabilities for $\delta_2=2\pi \times 5$ MHz, $2\pi \times 2.5$ MHz, $-2\pi \times 2.5$ MHz, and $-2\pi \times 5$ MHz, respectively.
   Calculations based on ideal situations are on the left and experimental data are on the right.
   All the graphs here share the same color bar.
}
\label{retprob}
\end{figure}

The return probabilities for $\delta_2=2\pi \times 5$ MHz, $2\pi \times 2.5$ MHz, $-2\pi \times 2.5$ MHz, and $-2\pi \times 5$ MHz are illustrated in Figs.~\ref{retprob}(a)-(d), respectively. The return probabilities for $\delta_2=2\pi \times 5$ MHz and $-2\pi \times 5$ MHz are also shown in Figs.~\ref{rate}(a) and 3(b) of the main text.

%The rate functions corresponding to these return probabilities are illustrated in Fig.~\ref{ratefunc}.
%The experimental values of rate functions here carry error bars, which represent $\pm1$ s.d.
%These error bars are calculated according to the protocol of error propagation, from the standard deviations of the return probabilities.
%Some error bars are extraordinary long due to the nearly vanishing return probabilities around the critical quasimomenta and critical times.

\begin{figure}
\includegraphics[width=0.979\columnwidth]{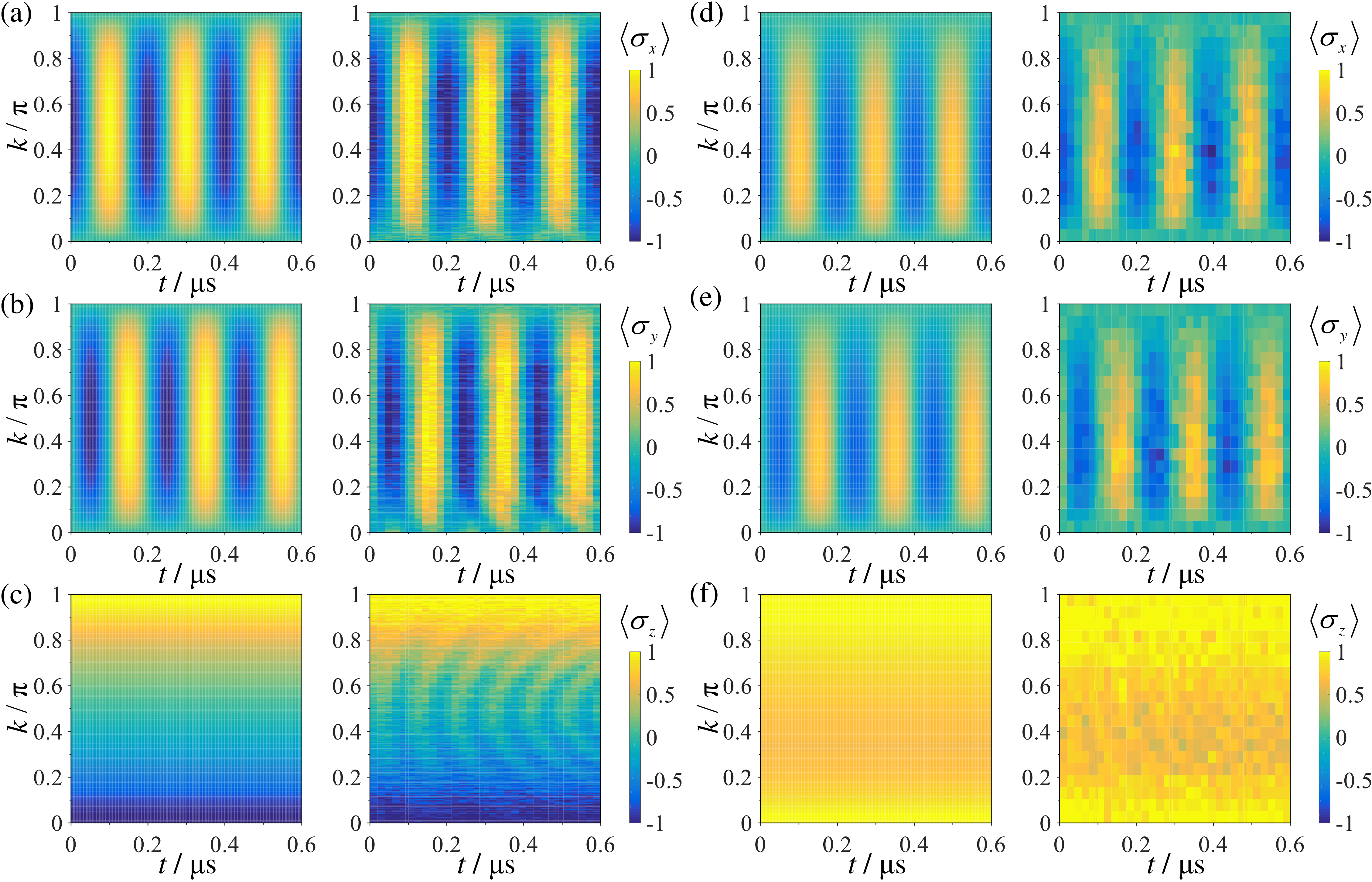}
\caption{Expectation values of $\sigma_x$, $\sigma_y$, and $\sigma_z$.
(a)-(c) Expectation values of $\sigma_x$, $\sigma_y$, and $\sigma_z$ as a function of the synthetic quasimomentum $k$ and the scaled time $\tau$ for $\delta_2=5$ MHz.
(d)-(f) Expectation values of $\sigma_x$, $\sigma_y$, and $\sigma_z$ as a function of the synthetic quasimomentum $k$ and the scaled time $\tau$ for $\delta_2=-5$ MHz.
Calculations based on ideal situations are on the left and experimental data are on the right.
}
\label{expectation}
\end{figure}

The geometric phase in Fig.~\ref{geometry}(a) and (b) of the main text are experimentally acquired from the expectation values of $\sigma_x$, $\sigma_y$, and $\sigma_z$.
These expectation values are illustrated in Fig.~\ref{expectation}.

%The dynamical topological invariants calculated from the geometric phases are shown in Fig.~\ref{topology}.
%Here the threshold for numerically discerning a discontinuity point is 1.6$\pi$.

%\begin{figure}\centering
%\includegraphics[width=1\columnwidth]{figs2}
%\caption{Rate functions.
%   (a)-(d) Rate functions for $\delta_2=2\pi \times 5$ MHz, $2\pi \times 2.5$ MHz, $-2\pi \times 2.5$ MHz, and $-2\pi \times 5$ MHz, respectively.
%   The black circles and the red curves represent experimental data and theoretical values, respectively.
%    Error bars represent $\pm1$ s.d.
%}
%\label{ratefunc}
%\end{figure}

%~\\~\\~\\~\\~\\~\\~\\~\\~\\~\\~\\~\\~\\~\\~\\~\\~\\~\\~\\~\\~\\~\\~\\~\\~\\~\\~\\~\\~\\~\\~\\~\\~\\~\\~\\~\\~\\~\\~\\~\\~\\~\\~\\~\\~\\~\\~\\~\\~\\~\\~\\~\\

%\begin{figure}\centering
%\includegraphics[width=1\columnwidth]{figs4}
%\caption{Dynamical topological order parameter.
%    (a)(b) Dynamical topological invariant for $\delta_2=5$ MHz and $\delta_2=-5$ MHz, respectively.
%    The black circles and the red lines represent experimental data and theoretical values, respectively.
%    Error bars represent $\pm1$ s.d.
%}
%\label{topology}
%\end{figure}
~\\~\\~\\~\\~\\~\\~\\~\\~\\~\\~\\~\\~\\~\\~\\~\\~\\~\\~\\~\\


\begin{thebibliography}{99}
\bibitem{noerrorbar}
The error bars of the experimental rate functions are not shown here because some bars are extroadanary long.
This is due to the nearly vanishing return probabilities around the critical quasimomenta and critical times.
\end{thebibliography}
\end{document}